\documentclass[%
 reprint,
nofootinbib,
 amsmath,amssymb,
 prl,
]{revtex4-2}

\usepackage{graphicx}
\usepackage{dcolumn}
\usepackage{bm}

\usepackage{tikz,xcolor,hyperref}
\hypersetup{
    colorlinks,
    linkcolor={red!80!black},
    citecolor={blue!50!black},
    urlcolor={blue!80!black}
}

\usepackage[
total={6.5in,9.75in}, top=0.9 in, left=0.9in, includefoot,
]{geometry}
\usepackage{color}

\begin{document}

\preprint{APS/123-QED}


\title{Elastic turbulence hides in the small scales of inertial polymeric turbulence}

\author{Piyush Garg}
\email{piyush.garg@oist.jp}
\affiliation{Complex Fluids and Flows Unit, Okinawa Institute of Science and Technology Graduate University, Okinawa 904-0495, Japan}
\author{Marco Edoardo Rosti}%
\email{marco.rosti@oist.jp}
\affiliation{Complex Fluids and Flows Unit, Okinawa Institute of Science and Technology Graduate University, Okinawa 904-0495, Japan}

\begin{abstract}
Gaining a fundamental understanding of turbulent flows of dilute polymer solutions has been a challenging and outstanding problem for a long time. In this letter, we examine homogeneous, isotropic polymeric turbulence at large Reynolds and Deborah numbers through direct numerical simulations. While at the largest scales of the flow inertial turbulence exists, we find that the flow is fundamentally altered from Newtonian turbulence below the Kolmogorov scale. We demonstrate that `Elastic Turbulence' exists at the smallest scales of polymeric turbulence by quantifying multiple statistical properties of the flow - energy spectrum and flux in Fourier space as well as the spatial statistics of the velocity field - the structure functions and kurtosis, and energy dissipation. Our results show the coexistence of two fundamentally distinct types of turbulence in polymeric fluids and point to the ubiquity of elastic turbulence, which was hitherto only known to exist for negligible inertia. 
\end{abstract}

\maketitle
Turbulent flows of dilute polymer solutions have been of long standing interest since the early work of Toms \cite{toms1949some}. In addition to the Reynolds number ($Re$), which is a measure of fluid inertia, polymeric fluids are characterized by the Deborah number ($De$) - a measure of the elasticity of the polymers \cite{bird1987dynamics, larson2013constitutive}. The interaction between inertia, elasticity and viscous dissipation means that polymeric fluids can exhibit a wide variety of turbulent flow behavior. The most well-known is that the addition of even a small concentration of polymers can dramatically modify wall-bounded turbulent flows, leading to reduced drag at large $Re$ and $De$ \cite{virk75, mung08, graham2014drag, benzi2018polymers}, as well as early transition to turbulence at moderate $Re$ \cite{samanta2013elasto, garg2018viscoelastic, dubief2023elasto}. The non-trivial nature of polymeric turbulence is not limited to wall bounded flows, as polymers affect both energy dissipation and associated flow structures in grid generated homogeneous, isotropic turbulence, as shown in several experiments \cite{barnard1969grid, mccomb1977effect, van1999decay, liberzon2006turbulent, perlekar2006manifestations, vonlanthen2013grid,  ouellette2009bulk, xi2013elastic} and simulations \cite{berti2006small, perlekar2010direct, cai2010dns, watanabe2013hybrid, watanabe2014power, valente2014effect, nguyen2016small}. Recently, a novel `elasto-inertial range' scaling was discovered in polymeric turbulence for large $Re$ and $De \sim$ O$(1)$, where the kinetic energy spectrum scales as $E \sim k^{-2.3}$ instead of the usual Kolmogorov scaling ($\sim k^{-5/3}$) \cite{zhang2021experimental, rosti2023large, chiarini2024extending}. Strikingly, polymeric fluids also exhibit a distinct spatio-temporally chaotic state at negligible $Re$ and large $De$, often termed as `Elastic turbulence' \cite{groisman2000elastic, larson2000turbulence, boffetta2003two, groisman2004elastic, burghelea2007elastic, steinberg2019scaling, steinberg2021elastic}. Unlike inertial turbulence, elastic turbulence involves a balance between the polymeric stresses and viscous dissipation, with negligible contribution from fluid inertia, and is characterized by a novel scaling of the fluid energy spectrum, $E(k) \sim k^{-4}$ \cite{steinberg2019scaling, steinberg2021elastic, lellep2024purely, singh2024intermittency, rota2024unified, singh2024interplay}. It has been assumed that elastic and inertial turbulence occur in different parameter regimes, with elastic turbulence only existing at low global $Re$, so inertia is negligible everywhere.

Polymers modify turbulent flows through a wide range of mechanics and at multiple length scales, a comprehensive understanding of which is still lacking after years of study. Here, we examine the small scales of homogeneous, isotropic turbulence of polymer solutions using direct numerical simulations (DNS). A few existing studies using DNS \cite{perlekar2010direct, watanabe2013hybrid, watanabe2014power, nguyen2016small} have suggested that polymers modify the fluid energy spectrum also in the dissipation range. One experiment, using PEO solution in grid turbulence, also observed a modified energy spectrum, but only transiently \cite{vonlanthen2013grid}. Despite these results, the exact nature of the flow at small scales in polymeric turbulence is largely unknown.

In this letter, we characterize the statistical properties of the flow at small scales in large $Re-De$ polymeric turbulence, and show that they are the same as those of elastic turbulence occurring at negligible $Re$. Our results thus reveal that elastic turbulence need not be confined to low global $Re$, as has been presumed so far since it's discovery \cite{groisman2000elastic, larson2000turbulence, steinberg2021elastic}. We indeed demonstrate that inertial and elastic turbulence simultaneously co-exist at the large and small scales in turbulent flows of polymers. 

\begin{figure}
		\includegraphics[scale=0.5]{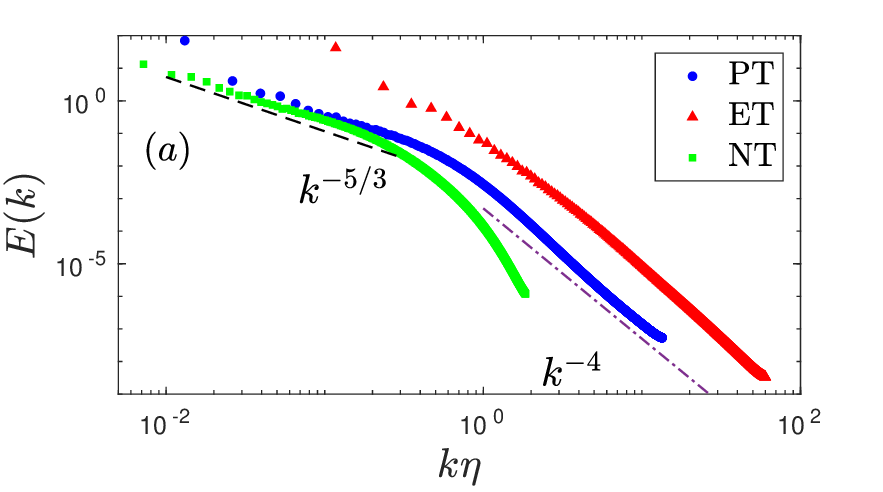}
		\includegraphics[scale=0.5]{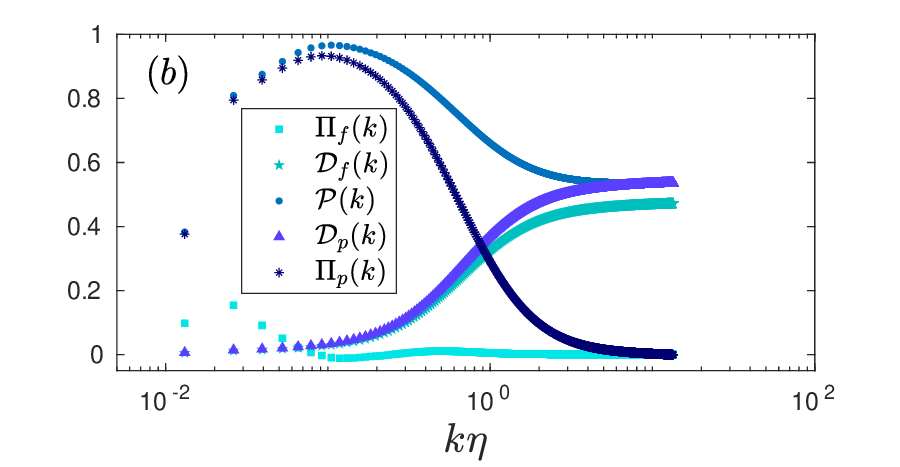}
        \includegraphics[scale=0.5]{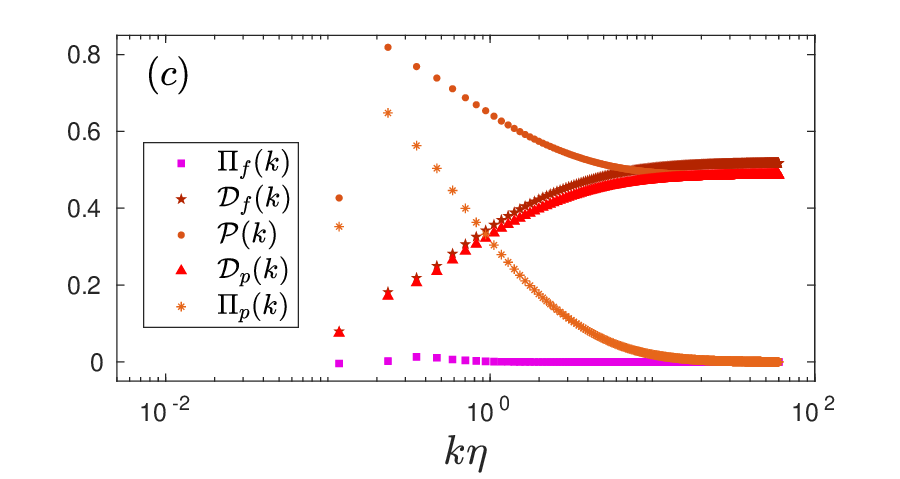}
	\caption{$(a)$ Energy spectrum $E(k)$ versus the non-dimensional wavenumber $k \eta$, for polymeric (PT), elastic (ET), and Newtonian turbulence (NT). The various energy fluxes in Fourier space versus $k \eta$ for $(b)$ polymeric and $(c)$ elastic turbulence.}
    \label{fig:spectra}
\end{figure}

The dynamics of polymeric fluids are described by the Navier-Stokes and continuity equation for the fluid velocity field $\boldsymbol{u}$, coupled to a constitutive model for the evolution of the polymer conformation tensor $\boldsymbol{C}$. We use the Oldroyd-B model, which assumes that the polymer molecules deform affinely with the flow, and has already been shown to successfully capture the fundamental physics of polymeric fluids \cite{bird1987dynamics, larson2013constitutive}. Thus, the governing equations are
\begin{equation}
\rho_f \left( \partial_t \boldsymbol{u} +  \left( \boldsymbol{u} \cdot \boldsymbol{\nabla} \right) \boldsymbol{u} \right)=-\boldsymbol{\nabla} p+\frac{\mu_p}{\tau_p} \boldsymbol{\nabla} \cdot \boldsymbol{C} + \mu_f \nabla^2 {\boldsymbol{u}} + \boldsymbol{F}, 
\nonumber
\end{equation}
\begin{equation}
\boldsymbol{\nabla} \cdot \boldsymbol{u} =0,
\nonumber
\end{equation}
\begin{equation}
\partial_t \boldsymbol{C} + \boldsymbol{u} \cdot \boldsymbol{\nabla} \boldsymbol{C} = \boldsymbol{C} \cdot \boldsymbol{\nabla}  \boldsymbol{u} + \boldsymbol{\nabla} \boldsymbol{u}^{T} \cdot \boldsymbol{C} - \frac{\boldsymbol{C}-\textbf{I}}{\tau_p},
\nonumber
\end{equation}
where $\rho_f$ is the fluid density, $\mu_f$ the fluid viscosity, $\mu_p$ the additional viscosity due to the polymers, $\tau_p$ the relaxation time of the polymer molecules, and the force $F$ is used to sustain the turbulence. See the Supplementary Materials for additional details of the numerical simulation \cite{fn}.  

In order to focus on the large $Re$-$De$ regime, we choose the Taylor Reynolds number $Re_\lambda \approx 240$ and the Deborah number $De = \tau_p/\tau_L \approx 9$, where $\tau_L$ is the turnover time of the largest eddy, with $\beta = \mu_s/(\mu_s+ \mu_p) = 0.9$. We term this parameter set, $Re_\lambda \approx 240, De \approx 9$ as Polymeric Turbulence (PT) in the following. We use a grid size of $2048^3$ grid points for PT; this large resolution has not been previously achieved in DNS at large $Re-De$ and is required to sufficiently resolve the statistics at small scales. Furthermore, we consider Elastic Turbulence (ET) and Newtonian Turbulence (NT) to compare, where we set $Re_\lambda \approx 40, De  \approx 9,$ $\beta  = 0.9 $ and $Re_\lambda \approx 460 , De = 0$, respectively, simulated with $1024^3$ grid points. To compare the different cases, we also define the Kolmogorov length scale, $\eta = \nu_f^3/\epsilon_f$, where $\nu_f = \mu_f/\rho_f$ and $\epsilon_f$ is the viscous energy dissipation by the solvent. Recall that in Newtonian turbulence, inertia and viscous dissipation dominate for $k \eta <1$ and $k \eta >1$, respectively \cite{frisch1995turbulence, davidson2015turbulence}.

First, we examine the fluid energy spectrum $E(k)$ in Fourier space versus $k \eta$, which shows two distinct scaling ranges in Fig. \ref{fig:spectra}$(a)$. For small wavenumbers ($k \eta < 1$), the energy spectrum $E(k)$ in PT scales as $\sim k^{-5/3}$ like in classical NT. Here, since we have $De \gg 1$ with $\beta =0.9$, following earlier work \cite{rosti2023large}, we recover the Kolmogorov scaling in the inertial range. At larger wavenumbers ($k\eta >1$), we see a separate power-law scaling of $E \sim k^{-4}$, differently from NT where the energy spectrum decays exponentially \cite{chen1993far, schumacher2007sub, khurshid2018energy, gorbunova2020analysis}. In PT we see no hint of an exponential scaling even at the largest wavenumbers resolved, with instead the observed power-law scaling corresponding to the one of ET \cite{groisman2000elastic, burghelea2007elastic, lellep2024purely, singh2024intermittency, steinberg2021elastic}. Remarkably, the Kolmogorov length scale $\eta$ continues to demarcate the two different regimes. The Supplementary Material \cite{fn} shows the kinetic energy spectrum in PT obtained using the FENE-P model for the polymers, where we once again find the $k^{-4}$ regime. The results presented here are therefore independent of the constitutive model used.  
  
\begin{figure}
		\includegraphics[scale=0.5]{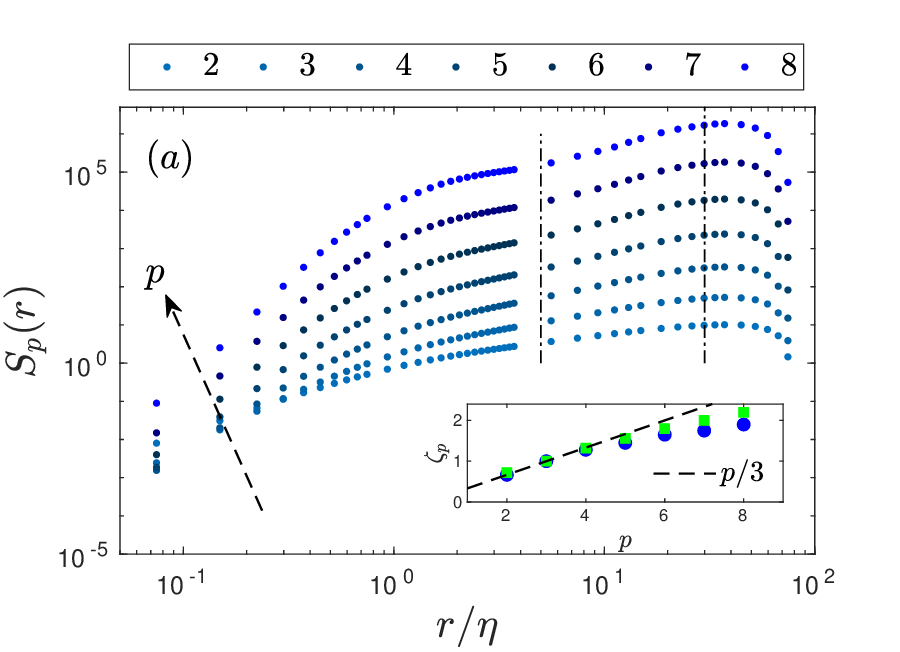}
		\includegraphics[scale=0.5]{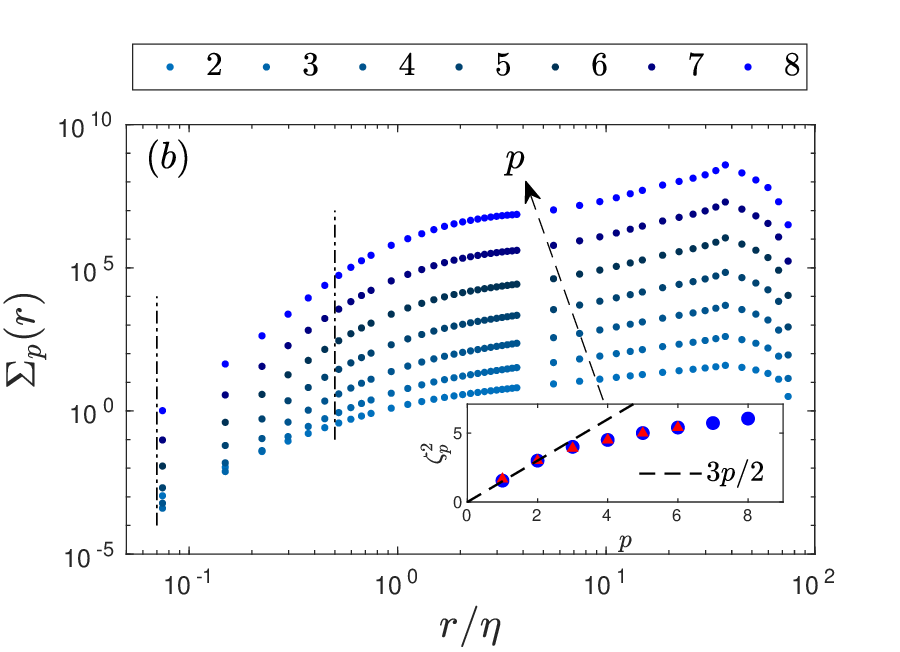}
	\caption{The $r$-variation of the structure functions $(a)$ $S_p$ and $(b)$ $\Sigma_p$, for $p$ from 2 to 8 for polymeric turbulence. The insets show the associated intermittency exponents in the inertial and `elastic-dissipation' ranges in $(a)$ and $(b)$, respectively (circles for PT), with the range where the scaling exponents are extracted marked in both figures. For comparison, we also show the results from \cite{iyer2020scaling} for inertial turbulence and \cite{singh2024intermittency} for elastic turbulence in the insets of $(a)$ and $(b)$ as squares and triangles, respectively.}
    \label{fig:str_fnc}
\end{figure}

To identify the dominant physical mechanism in each regime, we look at the variation of the energy flux across wavenumbers, shown in Fig. \ref{fig:spectra}$(b)$. In polymeric turbulence, there are three distinct ways for energy redistribution - inertial $\Pi_f$, viscous $D_f$, and polymeric $\mathcal{P}$ \cite{rosti2023large, singh2024interplay, valente2014effect}; see Supplementary Material for the derivation \cite{fn}. As expected, the inertial flux is only non-zero at the largest scales ($k \eta \ll 1$), while viscous dissipation is active at small scales ($k \eta \gg 1$). The polymeric flux has a non-trivial contribution across all wavenumbers, showing that the elastic stresses continue to play an important role even for $k \eta \gg 1$ \cite{Fn2}. $\mathcal{P}$ can further be partitioned into a dissipative ($D_p$) and non-dissipative ($\Pi_p$) contribution, where $D_p \rightarrow \mathcal{P}$ for $k \eta \gg 1$ \cite{rosti2023large}. With this decomposition, only the inertial ($\Pi_f$) and non-dissipative component of the polymeric ($\Pi_p$) flux contribute to $k \eta \ll 1$, while both $D_p$ and $D_f$ dominate for $k \eta \gg 1$. Thus, we have a non-trivial balance between energy transfer due to polymers and viscous dissipation at large wavenumbers, once again reminiscent of ET as shown in Fig. \ref{fig:spectra}$(c)$. The polymeric energy spectra for PT and ET also have the same scaling for $k\eta \gg 1$, see the Supplementary Material \cite{fn}. Furthermore, the kinetic energy spectrum in the small scales of PT ($E(k) \sim k^{-4}$) agrees well with the existing experimental results for ET at low $Re$, which find the exponent to range from $3.3$ to $3.6$ \cite{groisman2000elastic, burghelea2007elastic, steinberg2019scaling, steinberg2021elastic} as well as numerical simulations \cite{lellep2024purely, berti2008two, singh2024intermittency} and theoretical predictions for $E(k)$ \cite{fouxon2003spectra}. However, the theory predicts different polymeric energy spectra $E_p(k)$ \cite{fn}.

Thus, we have shown that the usual dissipation range is replaced by an `elastic-dissipation' range in polymeric turbulence. Both the energy spectrum and flux suggest that ET exists in this range, but this is not conclusive by itself. In the remainder of the paper, we crucially demonstrate that these small scales also share the same velocity field and dissipation statistics as ET. 

The spatial statistics of the velocity field in turbulence are revealed by the structure functions of two-point velocity differences $S_p(r) = \langle \delta u(r)^p \rangle$, with $\delta u(r) = \left( \boldsymbol{u}(\boldsymbol{x}+\boldsymbol{r}) - \boldsymbol{u}(\boldsymbol{x}) \right) \cdot \hat{\boldsymbol{r}}$. Kolmogorov originally conjectured that the structure functions in NT would follow a similarity scaling, $S_p \sim r^{\zeta_p}$, predicting $\zeta_p = p/3$ in the inertial range based on dimensional analysis \cite{frisch1995turbulence}. Fig. \ref{fig:str_fnc}$(a)$ suggests that there are two distinct scaling regimes for $S_p$ in PT: for $r/\eta > 1$, we indeed see Kolmogorov's prediction for $S_3 \sim r$, i.e., $\zeta_3 = 1$, while the scaling exponents $\zeta_p$ are known to differ from $p/3$ for larger $p$ due to the intermittency \cite{sreenivasan1997phenomenology, ishihara2009study, davidson2012ten, iyer2020scaling}. The intermittency exponents $\zeta_p$ can be estimated from the slope of $S_p$ versus $r$ in the inertial range, and are plotted in the inset of Fig. \ref{fig:str_fnc}(a), where $\zeta_p$ for PT are seen to be the same as in the inertial range of NT \cite{sreenivasan1997phenomenology, ishihara2009study, davidson2012ten, iyer2020scaling, rosti2023large}. Thus, at large scales the inertial range intermittency statistics are recovered in polymeric turbulence. Instead, $S_p \sim r^p$ for $r/ \eta < 1$, which follows from the trivial scaling of the velocity field $u \sim r$ for $r \ll 1$. Such analytic behavior of the velocity field is expected whenever the fluid energy spectrum decays faster than $k^{-3}$ \cite{frisch1995turbulence}, which holds for both Newtonian and elastic turbulence. Hence, the small $r$ behavior of $S_p(r)$ by itself is insufficiently revealing \cite{schumacher2007asymptotic, singh2024intermittency}.  

\begin{figure}
		\includegraphics[scale=0.5]{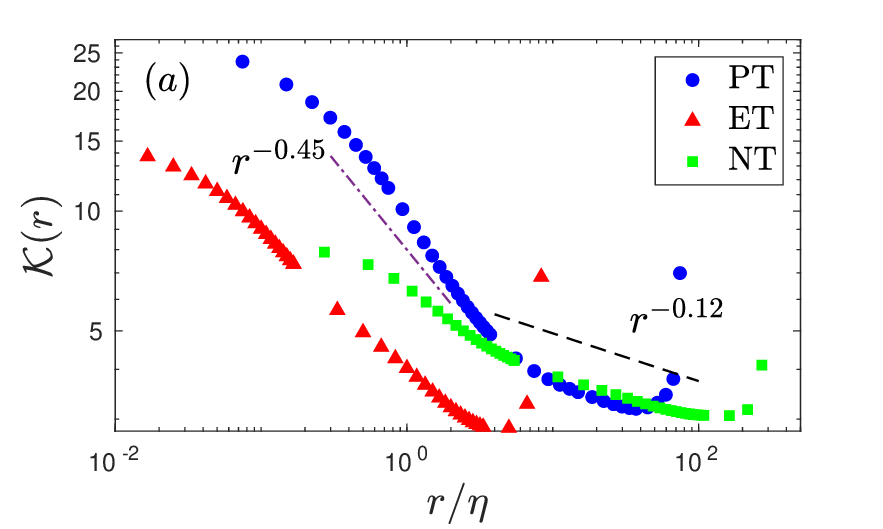}
        \includegraphics[scale=0.5]{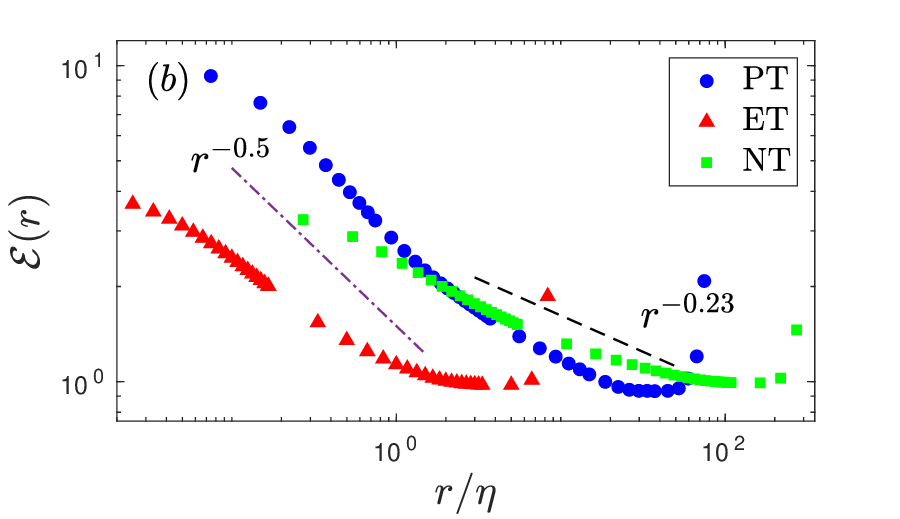}
	\caption{$(a)$ The kurtosis $\mathcal{K}(r) = \langle \delta u(r)^4 \rangle / \langle \delta u(r)^2 \rangle^2$ and $(b)$ the covariance of the fluid dissipation $\mathcal{E}(r)$, versus the non-dimensional distance $r/\eta$, for polymeric (PT), elastic (ET) and Newtonian (NT) turbulence.}
    \label{fig:kurtosis}
\end{figure}

To differentiate the dynamics at small scales, we look at the structure functions of the symmetric three-point velocity difference, $\Sigma_p(r) = \langle \delta^2 u(r)^p \rangle$, with $\delta^2 u(r) = \left( \boldsymbol{u}(\boldsymbol{x}+\boldsymbol{r}) - 2\boldsymbol{u}(\boldsymbol{x})+\boldsymbol{u}(\boldsymbol{x}-\boldsymbol{r}) \right) \cdot \hat{\boldsymbol{r}}$, since it was recently shown that $\Sigma_p$ reveals the intermittency in ET by removing the leading order analytic contribution  \cite{singh2024intermittency}. Fig. \ref{fig:str_fnc}$(b)$ shows that $\Sigma_p(r) \sim r^{\zeta_p^2}$ also has two distinct scaling regimes in PT, just like $S_p(r)$. In the inertial range, the velocity field is already rough at the leading order, so $\Sigma_p$ shows the same scaling behavior as $S_p$, i.e., $\zeta_p^2 \approx \zeta_p$ for $r/\eta>1$ (not shown). In contrast, for $r/\eta < 1$, $\Sigma_p(r)$ has a nontrivial scaling behavior, distinct from that of $S_p(r)$. Scaling analysis would suggest $\zeta_p^2 = 3/2$ for $r/\eta < 1$ \cite{singh2024intermittency}, but the exponents reveal intermittency corrections and saturate below $3p/2$ for large $p$ (see the inset of Fig. \ref{fig:str_fnc}$(b)$). More importantly, for $r/\eta<1$, the exponents $\zeta_p^2$ are the same as those discovered in ET \cite{singh2024intermittency}, and thus, the statistics of the velocity fluctuations at the small scales in polymeric turbulence, including the intermittency behavior, are exactly the same as elastic turbulence. To further confirm this, the Supplementary Material \cite{fn} shows the comparison of $\Sigma_2$ for PT and NT  for $r/\eta \ll 1$; for NT, $\Sigma_2(r) \sim r^4$, confirming the analyticity of the velocity field, while for PT $\Sigma_2(r) \sim r^3$.

\begin{figure}
		\includegraphics[scale=0.5]{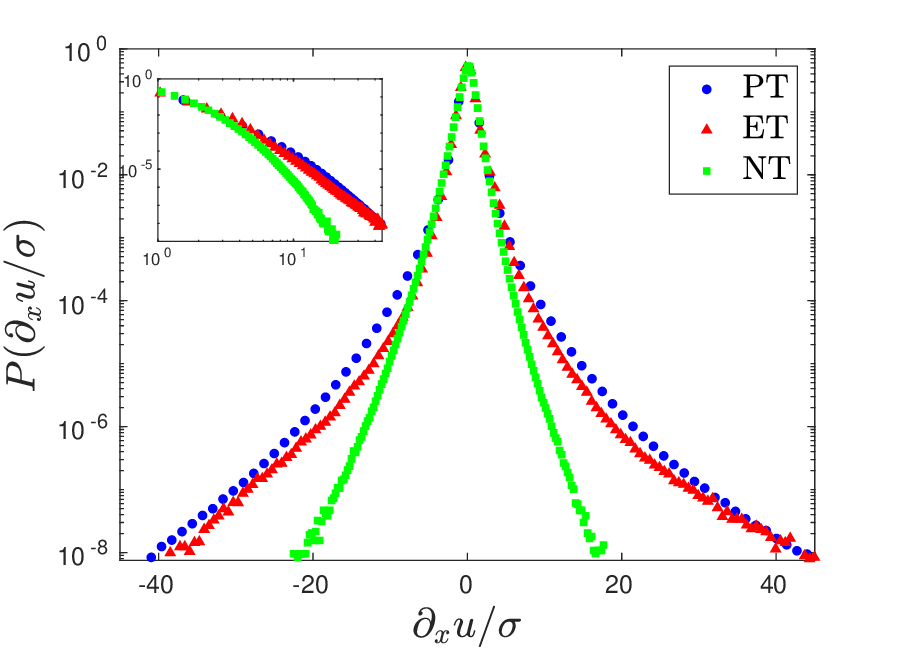}
	\caption{The probability distribution function (pdf) of the normalized, longitudinal velocity gradient ($\partial_x u/\sigma$) for PT, ET and NT, where $\sigma$ is the standard deviation; the inset shows the tails of the pdf on a log-log plot.}
    \label{fig:pdf}
\end{figure}

We look at the kurtosis $\mathcal{K}(r) = \langle \delta u(r)^4 \rangle / \langle \delta u(r)^2 \rangle^2$, a measure of the deviation from a Gaussian distribution  for which $\mathcal{K}=3$; a larger kurtosis implies elongated tails with increased contribution from rare but extreme events, and $\mathcal{K}$ is used as another measure of the intermittency of the velocity field in NT \cite{frisch1995turbulence, ishihara2009study, davidson2012ten}. Fig. \ref{fig:kurtosis}$(a)$ shows two distinct scaling regimes of $\mathcal{K}$ for PT: for $r/\eta > 1$, $\mathcal{K}$ shows the same scaling behavior as the inertial range of NT ($\mathcal{K} \sim r^{-0.12}$) \cite{ishihara2009study, davidson2012ten, iyer2020scaling}, while the scaling of the kurtosis is seen to be similar to ET for $r/\eta <1$ ($\mathcal{K} \sim r^{-0.45}$). Further, since the kurtosis at small scales in PT increases faster than in the dissipation regime of NT, the velocity field in the `elastic-dissipation' range of polymeric turbulence at large $Re-De$ is thus more intermittent than the usual dissipation range in NT.

\begin{figure}
		\includegraphics[scale=0.5]{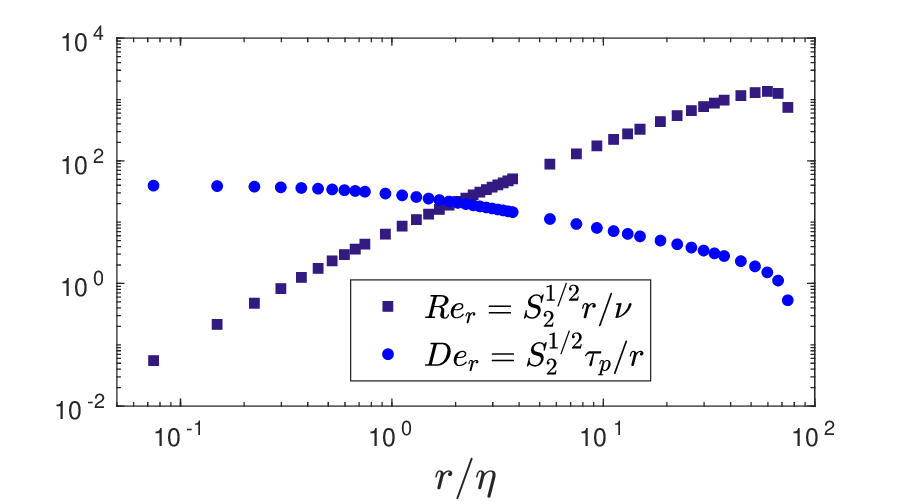}
	\caption{The variation of the local Reynolds ($Re_r$) and Deborah ($De_r$) number with $r/\eta$ for polymeric turbulence with $Re_\lambda \approx 240$ and $De \approx 9$.}
    \label{fig:local}
\end{figure}

Next, we examine the spatial statistics of the fluid dissipation $\epsilon_f = 2 \nu_f  \boldsymbol{S} \boldsymbol{:}  \boldsymbol{S} $, with $\boldsymbol{S} =1/2 (\nabla \boldsymbol{u}+ (\nabla \boldsymbol{u})^{T})$, using the spatial autocorrelation defined as $\mathcal{E}(r) = \langle \epsilon_f(x) \epsilon_f(x+r) \rangle / \langle \epsilon_f \rangle^2 $. In the inertial range of NT, $\mathcal{E} \sim r^{-\mu}$, where the exponent $\mu$ is a measure of the intermittent and multifractal nature of energy dissipation \cite{nelkin1981dissipation, cates1987spatial, sreenivasan1993update, meneveau1991multifractal, buaria2022intermittency}. In polymeric turbulence, $\mathcal{E}$ again has two distinct scaling ranges, as shown in Fig. \ref{fig:kurtosis}$(b)$: the inertial ($\mathcal{E} \sim r^{-0.23} $) and the `elastic-dissipation' ($\mathcal{E} \sim r^{-0.5} $) range, corresponding to Newtonian \cite{nelkin1981dissipation, sreenivasan1993update, buaria2022intermittency} and elastic turbulence, respectively. The increased scaling exponent at the smaller scales also implies that energy dissipation in polymeric turbulence is much more intermittent, in agreement with the results obtained using the kurtosis earlier. Finally, Fig. \ref{fig:pdf} compares the probability distribution function of the normalized, longitudinal velocity gradient in the three cases. The statistics of the velocity gradients independently characterize the small scales of turbulence and are important in determining the dynamics of sub-Kolmogorov scale particles as well as mixing \cite{sreenivasan1997phenomenology, ishihara2009study, davidson2012ten, schumacher2014small}. Once again, the spatial structure of the flow at the small scales in polymeric turbulence well corresponds to that of elastic turbulence.   

We have conclusively demonstrated that elastic turbulence exists for $r/\eta \ll 1$, while inertial turbulence occurs for $r/\eta \gg 1$ at large $Re_\lambda$ and $De$. We now investigate \textit{why} and \textit{when} does elastic turbulence exist at the small scales of large $Re_\lambda$.

At a given length scale $r$, polymers experience the shear-rate $\dot{\gamma}_r = (S_2)^{1/2}/r$, where $(S_2)^{1/2} = \delta u(r)$ is a scale-dependent measure of the fluid velocity. We can now define a local Deborah number $De_r = \tau_p (S_2)^{1/2}/r$, and, similarly, a scale dependent Reynolds number $Re_r = S_2^{1/2} r/\nu $ \cite{tabor1986cascade, de1986towards}. Fig. \ref{fig:local} shows the variation of $Re_r$ and $De_r$ with $r/\eta$. At large scales ($r/\eta > 1$), $Re_r \gg De_r $ and we see that inertia dominates. At small scales ($r/\eta < 1$), $De_r$ asymptotes to a constant ($\sim r^0$ since $S_2^{1/2} \sim r$), while $Re_r$ keeps decreasing with $r$($\sim r^2$); thus, as the inertial effects decrease at smaller scales, eventually we have a transition length scale where $Re_r$ becomes smaller than $De_r$. For ET to exist, the velocity fluctuations at small scales must be strong enough to sufficiently stretch the polymers, thus we can expect ET to appear when $De_r \gg Re_r$ and $De_r \gg 1$ for $r \to 0$ with $Re_r \ll 1$, exactly the parameter range at which elastic turbulence is expected. 

This suggests that elastic turbulence should exist if the local $De_r$ number is large. To verify this hypothesis, we examine the energy spectrum for $Re_\lambda \approx 240$ with additional Deborah numbers of $1/9, 1/3, 1$ and $3$, shown in the Supplementary Material \cite{fn}. For $De \approx 1/3$, $1$, and $3$, the local $De_r$ for $r/\eta \ll 1$ is greater than $1$, and indeed we observe the $k^{-4}$ spectrum and hence ET. On the contrary, for $De \approx 1/9$, $De_r$ for $r/\eta \ll 1$ is too small, and consequently ET does not exist and the energy spectrum recovers the usual exponential decay in the dissipation range.  

In this letter, we have shown that elastic turbulence is present in the small scales of turbulent flows of polymer solutions for a wide range of parameters -- for any (even large) $Re_\lambda$ whenever the local $De_r \geq 1$. We provide a crucial missing link in gaining a mechanistic understanding of polymeric turbulence by quantifying the flow at small scales, something that so far has lacked a quantitative understanding. Two competing theories have been proposed to explain drag reduction by polymer additives \cite{sreenivasan2000onset}: the viscous theory, where only the increased extensional viscosity due to polymers is important \cite{lumley1969drag, lumley1973drag, procaccia2008colloquium}, and the elastic theory, where elastic effects instead dominate \cite{tabor1986cascade, de1986towards, balkovsky2001turbulence, fouxon2003spectra}. Our results suggest that elastic effects non-trivially modify the small scales in homogeneous turbulence and thus cannot be neglected. However, we do not find the `elastic-waves' range predicted by later extensions of the elastic theory \cite{de1986towards, balkovsky2001turbulence, fouxon2003spectra}, where inertia and elasticity balance. Instead, we find a balance between elasticity and viscous dissipation, similar to ET at low global $Re$.

Finally, our results go beyond the expectations of existing theories by demonstrating that there is an exact correspondence between the statistics of PT at the small scales and ET. We thus expand the parameter range where ET is observed and show that it is quite ubiquitous in the flows of polymer solutions, a consistent theoretical description of which is still lacking \cite{steinberg2019scaling, steinberg2021elastic, datta2022perspectives, singh2024intermittency, lellep2024purely}. ET has also been used to engineer devices with enhanced mixing and reaction kinetics recently \cite{haward2021stagnation, browne2024harnessing}; therefore, our results also pose an intriguing question as to how ET at the small scale affects other global properties of the flow, such as mixing, in polymeric turbulence. Particularly an extension to wall-bounded flows could be relevant to polymeric drag reduction \cite{sreenivasan2000onset, mung08}. Finally, our results suggest future experimental measurements of structure functions and intermittency exponents for three-point velocity differences, extending recent work at large scales \cite{ouellette2009bulk, xi2013elastic, zhang2021experimental}.

\begin{acknowledgments}
\section*{Acknowledgments}
The research was supported by the Okinawa Institute of Science and Technology Graduate University (OIST) with subsidy funding to M.E.R. from the Cabinet Office, Government of Japan. M.E.R. also acknowledges funding from the Japan Society for the Promotion of Science (JSPS), grant 24K17210 and 24K00810. The authors acknowledge the computer time provided by the Scientific Computing \& Data Analysis section of the Core Facilities at OIST, and by HPCI, under the Research Project grants hp210269, hp220099, hp230018, hp250021, and hp250035.
\end{acknowledgments}

\bibliography{refer.bib}

\providecommand{\noopsort}[1]{}\providecommand{\singleletter}[1]{#1}%
\begin{thebibliography}{76}%
\makeatletter
\providecommand \@ifxundefined [1]{%
 \@ifx{#1\undefined}
}%
\providecommand \@ifnum [1]{%
 \ifnum #1\expandafter \@firstoftwo
 \else \expandafter \@secondoftwo
 \fi
}%
\providecommand \@ifx [1]{%
 \ifx #1\expandafter \@firstoftwo
 \else \expandafter \@secondoftwo
 \fi
}%
\providecommand \natexlab [1]{#1}%
\providecommand \enquote  [1]{``#1''}%
\providecommand \bibnamefont  [1]{#1}%
\providecommand \bibfnamefont [1]{#1}%
\providecommand \citenamefont [1]{#1}%
\providecommand \href@noop [0]{\@secondoftwo}%
\providecommand \href [0]{\begingroup \@sanitize@url \@href}%
\providecommand \@href[1]{\@@startlink{#1}\@@href}%
\providecommand \@@href[1]{\endgroup#1\@@endlink}%
\providecommand \@sanitize@url [0]{\catcode `\\12\catcode `\$12\catcode
  `\&12\catcode `\#12\catcode `\^12\catcode `\_12\catcode `\%12\relax}%
\providecommand \@@startlink[1]{}%
\providecommand \@@endlink[0]{}%
\providecommand \url  [0]{\begingroup\@sanitize@url \@url }%
\providecommand \@url [1]{\endgroup\@href {#1}{\urlprefix }}%
\providecommand \urlprefix  [0]{URL }%
\providecommand \Eprint [0]{\href }%
\providecommand \doibase [0]{https://doi.org/}%
\providecommand \selectlanguage [0]{\@gobble}%
\providecommand \bibinfo  [0]{\@secondoftwo}%
\providecommand \bibfield  [0]{\@secondoftwo}%
\providecommand \translation [1]{[#1]}%
\providecommand \BibitemOpen [0]{}%
\providecommand \bibitemStop [0]{}%
\providecommand \bibitemNoStop [0]{.\EOS\space}%
\providecommand \EOS [0]{\spacefactor3000\relax}%
\providecommand \BibitemShut  [1]{\csname bibitem#1\endcsname}%
\let\auto@bib@innerbib\@empty
\bibitem [{\citenamefont {Toms}(1949)}]{toms1949some}%
  \BibitemOpen
  \bibfield  {author} {\bibinfo {author} {\bibfnamefont {B.~A.}\ \bibnamefont
  {Toms}},\ }\bibfield  {title} {\bibinfo {title} {Some observations on the
  flow of linear polymersolutions through straight tubes at large reynolds
  numbers},\ }in\ \href@noop {} {\emph {\bibinfo {booktitle} {Proc. 1st Intl
  Congr. Rheol.}}},\ Vol.~\bibinfo {volume} {2}\ (\bibinfo {year} {1949})\ pp.\
  \bibinfo {pages} {135--141}\BibitemShut {NoStop}%
\bibitem [{\citenamefont {Bird}\ \emph {et~al.}(1987)\citenamefont {Bird},
  \citenamefont {Armstrong},\ and\ \citenamefont
  {Hassager}}]{bird1987dynamics}%
  \BibitemOpen
  \bibfield  {author} {\bibinfo {author} {\bibfnamefont {R.~B.}\ \bibnamefont
  {Bird}}, \bibinfo {author} {\bibfnamefont {R.~C.}\ \bibnamefont
  {Armstrong}},\ and\ \bibinfo {author} {\bibfnamefont {O.}~\bibnamefont
  {Hassager}},\ }\href@noop {} {\emph {\bibinfo {title} {Dynamics of polymeric
  liquids. Vol. 1: Fluid mechanics}}}\ (\bibinfo  {publisher} {John Wiley and
  Sons Inc., New York, NY},\ \bibinfo {year} {1987})\BibitemShut {NoStop}%
\bibitem [{\citenamefont {Larson}(2013)}]{larson2013constitutive}%
  \BibitemOpen
  \bibfield  {author} {\bibinfo {author} {\bibfnamefont {R.~G.}\ \bibnamefont
  {Larson}},\ }\href@noop {} {\emph {\bibinfo {title} {Constitutive equations
  for polymer melts and solutions: Butterworths series in chemical
  engineering}}}\ (\bibinfo  {publisher} {Butterworth-Heinemann},\ \bibinfo
  {year} {2013})\BibitemShut {NoStop}%
\bibitem [{\citenamefont {Virk}(1975)}]{virk75}%
  \BibitemOpen
  \bibfield  {author} {\bibinfo {author} {\bibfnamefont {P.}~\bibnamefont
  {Virk}},\ }\bibfield  {title} {\bibinfo {title} {Drag reduction
  fundamentals},\ }\href@noop {} {\bibfield  {journal} {\bibinfo  {journal}
  {AIChE J.}\ }\textbf {\bibinfo {volume} {21}},\ \bibinfo {pages} {625}
  (\bibinfo {year} {1975})}\BibitemShut {NoStop}%
\bibitem [{\citenamefont {White}\ and\ \citenamefont {Mungal}(2008)}]{mung08}%
  \BibitemOpen
  \bibfield  {author} {\bibinfo {author} {\bibfnamefont {C.}~\bibnamefont
  {White}}\ and\ \bibinfo {author} {\bibfnamefont {M.}~\bibnamefont {Mungal}},\
  }\bibfield  {title} {\bibinfo {title} {Mechanics and prediction of turbulent
  drag reduction with polymer additives.},\ }\href@noop {} {\bibfield
  {journal} {\bibinfo  {journal} {Annu. Rev. Fluid Mech.}\ }\textbf {\bibinfo
  {volume} {40}},\ \bibinfo {pages} {235} (\bibinfo {year} {2008})}\BibitemShut
  {NoStop}%
\bibitem [{\citenamefont {Graham}(2014)}]{graham2014drag}%
  \BibitemOpen
  \bibfield  {author} {\bibinfo {author} {\bibfnamefont {M.~D.}\ \bibnamefont
  {Graham}},\ }\bibfield  {title} {\bibinfo {title} {Drag reduction and the
  dynamics of turbulence in simple and complex fluids},\ }\href@noop {}
  {\bibfield  {journal} {\bibinfo  {journal} {Physics of Fluids}\ }\textbf
  {\bibinfo {volume} {26}} (\bibinfo {year} {2014})}\BibitemShut {NoStop}%
\bibitem [{\citenamefont {Benzi}\ and\ \citenamefont
  {Ching}(2018)}]{benzi2018polymers}%
  \BibitemOpen
  \bibfield  {author} {\bibinfo {author} {\bibfnamefont {R.}~\bibnamefont
  {Benzi}}\ and\ \bibinfo {author} {\bibfnamefont {E.~S.}\ \bibnamefont
  {Ching}},\ }\bibfield  {title} {\bibinfo {title} {Polymers in fluid flows},\
  }\href@noop {} {\bibfield  {journal} {\bibinfo  {journal} {Annual Review of
  Condensed Matter Physics}\ }\textbf {\bibinfo {volume} {9}},\ \bibinfo
  {pages} {163} (\bibinfo {year} {2018})}\BibitemShut {NoStop}%
\bibitem [{\citenamefont {Samanta}\ \emph {et~al.}(2013)\citenamefont
  {Samanta}, \citenamefont {Dubief}, \citenamefont {Holzner}, \citenamefont
  {Sch{\"a}fer}, \citenamefont {Morozov}, \citenamefont {Wagner},\ and\
  \citenamefont {Hof}}]{samanta2013elasto}%
  \BibitemOpen
  \bibfield  {author} {\bibinfo {author} {\bibfnamefont {D.}~\bibnamefont
  {Samanta}}, \bibinfo {author} {\bibfnamefont {Y.}~\bibnamefont {Dubief}},
  \bibinfo {author} {\bibfnamefont {M.}~\bibnamefont {Holzner}}, \bibinfo
  {author} {\bibfnamefont {C.}~\bibnamefont {Sch{\"a}fer}}, \bibinfo {author}
  {\bibfnamefont {A.~N.}\ \bibnamefont {Morozov}}, \bibinfo {author}
  {\bibfnamefont {C.}~\bibnamefont {Wagner}},\ and\ \bibinfo {author}
  {\bibfnamefont {B.}~\bibnamefont {Hof}},\ }\bibfield  {title} {\bibinfo
  {title} {Elasto-inertial turbulence},\ }\href@noop {} {\bibfield  {journal}
  {\bibinfo  {journal} {Proceedings of the National Academy of Sciences}\
  }\textbf {\bibinfo {volume} {110}},\ \bibinfo {pages} {10557} (\bibinfo
  {year} {2013})}\BibitemShut {NoStop}%
\bibitem [{\citenamefont {Garg}\ \emph {et~al.}(2018)\citenamefont {Garg},
  \citenamefont {Chaudhary}, \citenamefont {Khalid}, \citenamefont {Shankar},\
  and\ \citenamefont {Subramanian}}]{garg2018viscoelastic}%
  \BibitemOpen
  \bibfield  {author} {\bibinfo {author} {\bibfnamefont {P.}~\bibnamefont
  {Garg}}, \bibinfo {author} {\bibfnamefont {I.}~\bibnamefont {Chaudhary}},
  \bibinfo {author} {\bibfnamefont {M.}~\bibnamefont {Khalid}}, \bibinfo
  {author} {\bibfnamefont {V.}~\bibnamefont {Shankar}},\ and\ \bibinfo {author}
  {\bibfnamefont {G.}~\bibnamefont {Subramanian}},\ }\bibfield  {title}
  {\bibinfo {title} {Viscoelastic pipe flow is linearly unstable},\ }\href@noop
  {} {\bibfield  {journal} {\bibinfo  {journal} {Physical review letters}\
  }\textbf {\bibinfo {volume} {121}},\ \bibinfo {pages} {024502} (\bibinfo
  {year} {2018})}\BibitemShut {NoStop}%
\bibitem [{\citenamefont {Dubief}\ \emph {et~al.}(2023)\citenamefont {Dubief},
  \citenamefont {Terrapon},\ and\ \citenamefont {Hof}}]{dubief2023elasto}%
  \BibitemOpen
  \bibfield  {author} {\bibinfo {author} {\bibfnamefont {Y.}~\bibnamefont
  {Dubief}}, \bibinfo {author} {\bibfnamefont {V.~E.}\ \bibnamefont
  {Terrapon}},\ and\ \bibinfo {author} {\bibfnamefont {B.}~\bibnamefont
  {Hof}},\ }\bibfield  {title} {\bibinfo {title} {Elasto-inertial turbulence},\
  }\href@noop {} {\bibfield  {journal} {\bibinfo  {journal} {Annual Review of
  Fluid Mechanics}\ }\textbf {\bibinfo {volume} {55}},\ \bibinfo {pages} {675}
  (\bibinfo {year} {2023})}\BibitemShut {NoStop}%
\bibitem [{\citenamefont {Barnard}\ and\ \citenamefont
  {Sellin}(1969)}]{barnard1969grid}%
  \BibitemOpen
  \bibfield  {author} {\bibinfo {author} {\bibfnamefont {B.}~\bibnamefont
  {Barnard}}\ and\ \bibinfo {author} {\bibfnamefont {R.}~\bibnamefont
  {Sellin}},\ }\bibfield  {title} {\bibinfo {title} {Grid turbulence in dilute
  polymer solutions},\ }\href@noop {} {\bibfield  {journal} {\bibinfo
  {journal} {Nature}\ }\textbf {\bibinfo {volume} {222}},\ \bibinfo {pages}
  {1160} (\bibinfo {year} {1969})}\BibitemShut {NoStop}%
\bibitem [{\citenamefont {McComb}\ \emph {et~al.}(1977)\citenamefont {McComb},
  \citenamefont {Allan},\ and\ \citenamefont {Greated}}]{mccomb1977effect}%
  \BibitemOpen
  \bibfield  {author} {\bibinfo {author} {\bibfnamefont {W.}~\bibnamefont
  {McComb}}, \bibinfo {author} {\bibfnamefont {J.}~\bibnamefont {Allan}},\ and\
  \bibinfo {author} {\bibfnamefont {C.}~\bibnamefont {Greated}},\ }\bibfield
  {title} {\bibinfo {title} {Effect of polymer additives on the small-scale
  structure of grid-generated turbulence},\ }\href@noop {} {\bibfield
  {journal} {\bibinfo  {journal} {The Physics of Fluids}\ }\textbf {\bibinfo
  {volume} {20}},\ \bibinfo {pages} {873} (\bibinfo {year} {1977})}\BibitemShut
  {NoStop}%
\bibitem [{\citenamefont {van Doorn}\ \emph {et~al.}(1999)\citenamefont {van
  Doorn}, \citenamefont {White},\ and\ \citenamefont
  {Sreenivasan}}]{van1999decay}%
  \BibitemOpen
  \bibfield  {author} {\bibinfo {author} {\bibfnamefont {E.}~\bibnamefont {van
  Doorn}}, \bibinfo {author} {\bibfnamefont {C.~M.}\ \bibnamefont {White}},\
  and\ \bibinfo {author} {\bibfnamefont {K.}~\bibnamefont {Sreenivasan}},\
  }\bibfield  {title} {\bibinfo {title} {The decay of grid turbulence in
  polymer and surfactant solutions},\ }\href@noop {} {\bibfield  {journal}
  {\bibinfo  {journal} {Physics of Fluids}\ }\textbf {\bibinfo {volume} {11}},\
  \bibinfo {pages} {2387} (\bibinfo {year} {1999})}\BibitemShut {NoStop}%
\bibitem [{\citenamefont {Liberzon}\ \emph {et~al.}(2006)\citenamefont
  {Liberzon}, \citenamefont {Guala}, \citenamefont {Kinzelbach},\ and\
  \citenamefont {Tsinober}}]{liberzon2006turbulent}%
  \BibitemOpen
  \bibfield  {author} {\bibinfo {author} {\bibfnamefont {A.}~\bibnamefont
  {Liberzon}}, \bibinfo {author} {\bibfnamefont {M.}~\bibnamefont {Guala}},
  \bibinfo {author} {\bibfnamefont {W.}~\bibnamefont {Kinzelbach}},\ and\
  \bibinfo {author} {\bibfnamefont {A.}~\bibnamefont {Tsinober}},\ }\bibfield
  {title} {\bibinfo {title} {On turbulent kinetic energy production and
  dissipation in dilute polymer solutions},\ }\href@noop {} {\bibfield
  {journal} {\bibinfo  {journal} {Physics of Fluids}\ }\textbf {\bibinfo
  {volume} {18}} (\bibinfo {year} {2006})}\BibitemShut {NoStop}%
\bibitem [{\citenamefont {Perlekar}\ \emph {et~al.}(2006)\citenamefont
  {Perlekar}, \citenamefont {Mitra},\ and\ \citenamefont
  {Pandit}}]{perlekar2006manifestations}%
  \BibitemOpen
  \bibfield  {author} {\bibinfo {author} {\bibfnamefont {P.}~\bibnamefont
  {Perlekar}}, \bibinfo {author} {\bibfnamefont {D.}~\bibnamefont {Mitra}},\
  and\ \bibinfo {author} {\bibfnamefont {R.}~\bibnamefont {Pandit}},\
  }\bibfield  {title} {\bibinfo {title} {Manifestations of drag reduction by
  polymer additives in decaying, homogeneous, isotropic turbulence},\
  }\href@noop {} {\bibfield  {journal} {\bibinfo  {journal} {Physical review
  letters}\ }\textbf {\bibinfo {volume} {97}},\ \bibinfo {pages} {264501}
  (\bibinfo {year} {2006})}\BibitemShut {NoStop}%
\bibitem [{\citenamefont {Vonlanthen}\ and\ \citenamefont
  {Monkewitz}(2013)}]{vonlanthen2013grid}%
  \BibitemOpen
  \bibfield  {author} {\bibinfo {author} {\bibfnamefont {R.}~\bibnamefont
  {Vonlanthen}}\ and\ \bibinfo {author} {\bibfnamefont {P.~A.}\ \bibnamefont
  {Monkewitz}},\ }\bibfield  {title} {\bibinfo {title} {Grid turbulence in
  dilute polymer solutions: Peo in water},\ }\href@noop {} {\bibfield
  {journal} {\bibinfo  {journal} {Journal of Fluid Mechanics}\ }\textbf
  {\bibinfo {volume} {730}},\ \bibinfo {pages} {76} (\bibinfo {year}
  {2013})}\BibitemShut {NoStop}%
\bibitem [{\citenamefont {Ouellette}\ \emph {et~al.}(2009)\citenamefont
  {Ouellette}, \citenamefont {Xu},\ and\ \citenamefont
  {Bodenschatz}}]{ouellette2009bulk}%
  \BibitemOpen
  \bibfield  {author} {\bibinfo {author} {\bibfnamefont {N.~T.}\ \bibnamefont
  {Ouellette}}, \bibinfo {author} {\bibfnamefont {H.}~\bibnamefont {Xu}},\ and\
  \bibinfo {author} {\bibfnamefont {E.}~\bibnamefont {Bodenschatz}},\
  }\bibfield  {title} {\bibinfo {title} {Bulk turbulence in dilute polymer
  solutions},\ }\href@noop {} {\bibfield  {journal} {\bibinfo  {journal}
  {Journal of fluid mechanics}\ }\textbf {\bibinfo {volume} {629}},\ \bibinfo
  {pages} {375} (\bibinfo {year} {2009})}\BibitemShut {NoStop}%
\bibitem [{\citenamefont {Xi}\ \emph {et~al.}(2013)\citenamefont {Xi},
  \citenamefont {Bodenschatz},\ and\ \citenamefont {Xu}}]{xi2013elastic}%
  \BibitemOpen
  \bibfield  {author} {\bibinfo {author} {\bibfnamefont {H.-D.}\ \bibnamefont
  {Xi}}, \bibinfo {author} {\bibfnamefont {E.}~\bibnamefont {Bodenschatz}},\
  and\ \bibinfo {author} {\bibfnamefont {H.}~\bibnamefont {Xu}},\ }\bibfield
  {title} {\bibinfo {title} {Elastic energy flux by flexible polymers in fluid
  turbulence},\ }\href@noop {} {\bibfield  {journal} {\bibinfo  {journal}
  {Physical review letters}\ }\textbf {\bibinfo {volume} {111}},\ \bibinfo
  {pages} {024501} (\bibinfo {year} {2013})}\BibitemShut {NoStop}%
\bibitem [{\citenamefont {Berti}\ \emph {et~al.}(2006)\citenamefont {Berti},
  \citenamefont {Bistagnino}, \citenamefont {Boffetta}, \citenamefont
  {Celani},\ and\ \citenamefont {Musacchio}}]{berti2006small}%
  \BibitemOpen
  \bibfield  {author} {\bibinfo {author} {\bibfnamefont {S.}~\bibnamefont
  {Berti}}, \bibinfo {author} {\bibfnamefont {A.}~\bibnamefont {Bistagnino}},
  \bibinfo {author} {\bibfnamefont {G.}~\bibnamefont {Boffetta}}, \bibinfo
  {author} {\bibfnamefont {A.}~\bibnamefont {Celani}},\ and\ \bibinfo {author}
  {\bibfnamefont {S.}~\bibnamefont {Musacchio}},\ }\bibfield  {title} {\bibinfo
  {title} {Small-scale statistics of viscoelastic turbulence},\ }\href@noop {}
  {\bibfield  {journal} {\bibinfo  {journal} {Europhysics Letters}\ }\textbf
  {\bibinfo {volume} {76}},\ \bibinfo {pages} {63} (\bibinfo {year}
  {2006})}\BibitemShut {NoStop}%
\bibitem [{\citenamefont {Perlekar}\ \emph {et~al.}(2010)\citenamefont
  {Perlekar}, \citenamefont {Mitra},\ and\ \citenamefont
  {Pandit}}]{perlekar2010direct}%
  \BibitemOpen
  \bibfield  {author} {\bibinfo {author} {\bibfnamefont {P.}~\bibnamefont
  {Perlekar}}, \bibinfo {author} {\bibfnamefont {D.}~\bibnamefont {Mitra}},\
  and\ \bibinfo {author} {\bibfnamefont {R.}~\bibnamefont {Pandit}},\
  }\bibfield  {title} {\bibinfo {title} {Direct numerical simulations of
  statistically steady, homogeneous, isotropic fluid turbulence with polymer
  additives},\ }\href@noop {} {\bibfield  {journal} {\bibinfo  {journal}
  {Physical Review E—Statistical, Nonlinear, and Soft Matter Physics}\
  }\textbf {\bibinfo {volume} {82}},\ \bibinfo {pages} {066313} (\bibinfo
  {year} {2010})}\BibitemShut {NoStop}%
\bibitem [{\citenamefont {Cai}\ \emph {et~al.}(2010)\citenamefont {Cai},
  \citenamefont {Li},\ and\ \citenamefont {Zhang}}]{cai2010dns}%
  \BibitemOpen
  \bibfield  {author} {\bibinfo {author} {\bibfnamefont {W.-H.}\ \bibnamefont
  {Cai}}, \bibinfo {author} {\bibfnamefont {F.-C.}\ \bibnamefont {Li}},\ and\
  \bibinfo {author} {\bibfnamefont {H.-N.}\ \bibnamefont {Zhang}},\ }\bibfield
  {title} {\bibinfo {title} {Dns study of decaying homogeneous isotropic
  turbulence with polymer additives},\ }\href@noop {} {\bibfield  {journal}
  {\bibinfo  {journal} {Journal of Fluid Mechanics}\ }\textbf {\bibinfo
  {volume} {665}},\ \bibinfo {pages} {334} (\bibinfo {year}
  {2010})}\BibitemShut {NoStop}%
\bibitem [{\citenamefont {Watanabe}\ and\ \citenamefont
  {Gotoh}(2013)}]{watanabe2013hybrid}%
  \BibitemOpen
  \bibfield  {author} {\bibinfo {author} {\bibfnamefont {T.}~\bibnamefont
  {Watanabe}}\ and\ \bibinfo {author} {\bibfnamefont {T.}~\bibnamefont
  {Gotoh}},\ }\bibfield  {title} {\bibinfo {title} {Hybrid eulerian--lagrangian
  simulations for polymer--turbulence interactions},\ }\href@noop {} {\bibfield
   {journal} {\bibinfo  {journal} {Journal of Fluid Mechanics}\ }\textbf
  {\bibinfo {volume} {717}},\ \bibinfo {pages} {535} (\bibinfo {year}
  {2013})}\BibitemShut {NoStop}%
\bibitem [{\citenamefont {Watanabe}\ and\ \citenamefont
  {Gotoh}(2014)}]{watanabe2014power}%
  \BibitemOpen
  \bibfield  {author} {\bibinfo {author} {\bibfnamefont {T.}~\bibnamefont
  {Watanabe}}\ and\ \bibinfo {author} {\bibfnamefont {T.}~\bibnamefont
  {Gotoh}},\ }\bibfield  {title} {\bibinfo {title} {Power-law spectra formed by
  stretching polymers in decaying isotropic turbulence},\ }\href@noop {}
  {\bibfield  {journal} {\bibinfo  {journal} {Physics of Fluids}\ }\textbf
  {\bibinfo {volume} {26}} (\bibinfo {year} {2014})}\BibitemShut {NoStop}%
\bibitem [{\citenamefont {Valente}\ \emph {et~al.}(2014)\citenamefont
  {Valente}, \citenamefont {Da~Silva},\ and\ \citenamefont
  {Pinho}}]{valente2014effect}%
  \BibitemOpen
  \bibfield  {author} {\bibinfo {author} {\bibfnamefont {P.}~\bibnamefont
  {Valente}}, \bibinfo {author} {\bibfnamefont {C.}~\bibnamefont {Da~Silva}},\
  and\ \bibinfo {author} {\bibfnamefont {F.}~\bibnamefont {Pinho}},\ }\bibfield
   {title} {\bibinfo {title} {The effect of viscoelasticity on the turbulent
  kinetic energy cascade},\ }\href@noop {} {\bibfield  {journal} {\bibinfo
  {journal} {Journal of fluid mechanics}\ }\textbf {\bibinfo {volume} {760}},\
  \bibinfo {pages} {39} (\bibinfo {year} {2014})}\BibitemShut {NoStop}%
\bibitem [{\citenamefont {Nguyen}\ \emph {et~al.}(2016)\citenamefont {Nguyen},
  \citenamefont {Delache}, \citenamefont {Simo{\"e}ns}, \citenamefont {Bos},\
  and\ \citenamefont {El~Hajem}}]{nguyen2016small}%
  \BibitemOpen
  \bibfield  {author} {\bibinfo {author} {\bibfnamefont {M.~Q.}\ \bibnamefont
  {Nguyen}}, \bibinfo {author} {\bibfnamefont {A.}~\bibnamefont {Delache}},
  \bibinfo {author} {\bibfnamefont {S.}~\bibnamefont {Simo{\"e}ns}}, \bibinfo
  {author} {\bibfnamefont {W.~J.}\ \bibnamefont {Bos}},\ and\ \bibinfo {author}
  {\bibfnamefont {M.}~\bibnamefont {El~Hajem}},\ }\bibfield  {title} {\bibinfo
  {title} {Small scale dynamics of isotropic viscoelastic turbulence},\
  }\href@noop {} {\bibfield  {journal} {\bibinfo  {journal} {Physical Review
  Fluids}\ }\textbf {\bibinfo {volume} {1}},\ \bibinfo {pages} {083301}
  (\bibinfo {year} {2016})}\BibitemShut {NoStop}%
\bibitem [{\citenamefont {Zhang}\ \emph {et~al.}(2021)\citenamefont {Zhang},
  \citenamefont {Bodenschatz}, \citenamefont {Xu},\ and\ \citenamefont
  {Xi}}]{zhang2021experimental}%
  \BibitemOpen
  \bibfield  {author} {\bibinfo {author} {\bibfnamefont {Y.-B.}\ \bibnamefont
  {Zhang}}, \bibinfo {author} {\bibfnamefont {E.}~\bibnamefont {Bodenschatz}},
  \bibinfo {author} {\bibfnamefont {H.}~\bibnamefont {Xu}},\ and\ \bibinfo
  {author} {\bibfnamefont {H.-D.}\ \bibnamefont {Xi}},\ }\bibfield  {title}
  {\bibinfo {title} {Experimental observation of the elastic range scaling in
  turbulent flow with polymer additives},\ }\href@noop {} {\bibfield  {journal}
  {\bibinfo  {journal} {Science Advances}\ }\textbf {\bibinfo {volume} {7}},\
  \bibinfo {pages} {eabd3525} (\bibinfo {year} {2021})}\BibitemShut {NoStop}%
\bibitem [{\citenamefont {Rosti}\ \emph {et~al.}(2023)\citenamefont {Rosti},
  \citenamefont {Perlekar},\ and\ \citenamefont {Mitra}}]{rosti2023large}%
  \BibitemOpen
  \bibfield  {author} {\bibinfo {author} {\bibfnamefont {M.~E.}\ \bibnamefont
  {Rosti}}, \bibinfo {author} {\bibfnamefont {P.}~\bibnamefont {Perlekar}},\
  and\ \bibinfo {author} {\bibfnamefont {D.}~\bibnamefont {Mitra}},\ }\bibfield
   {title} {\bibinfo {title} {Large is different: nonmonotonic behavior of
  elastic range scaling in polymeric turbulence at large reynolds and deborah
  numbers},\ }\href@noop {} {\bibfield  {journal} {\bibinfo  {journal} {Science
  Advances}\ }\textbf {\bibinfo {volume} {9}},\ \bibinfo {pages} {eadd3831}
  (\bibinfo {year} {2023})}\BibitemShut {NoStop}%
\bibitem [{\citenamefont {Chiarini}\ \emph {et~al.}(2025)\citenamefont
  {Chiarini}, \citenamefont {Singh},\ and\ \citenamefont
  {Rosti}}]{chiarini2024extending}%
  \BibitemOpen
  \bibfield  {author} {\bibinfo {author} {\bibfnamefont {A.}~\bibnamefont
  {Chiarini}}, \bibinfo {author} {\bibfnamefont {R.~K.}\ \bibnamefont
  {Singh}},\ and\ \bibinfo {author} {\bibfnamefont {M.~E.}\ \bibnamefont
  {Rosti}},\ }\bibfield  {title} {\bibinfo {title} {Extending kolmogorov theory
  to polymeric turbulence},\ }\href@noop {} {\bibfield  {journal} {\bibinfo
  {journal} {Physical Review Research}\ }\textbf {\bibinfo {volume} {7}},\
  \bibinfo {pages} {L022043} (\bibinfo {year} {2025})}\BibitemShut {NoStop}%
\bibitem [{\citenamefont {Groisman}\ and\ \citenamefont
  {Steinberg}(2000)}]{groisman2000elastic}%
  \BibitemOpen
  \bibfield  {author} {\bibinfo {author} {\bibfnamefont {A.}~\bibnamefont
  {Groisman}}\ and\ \bibinfo {author} {\bibfnamefont {V.}~\bibnamefont
  {Steinberg}},\ }\bibfield  {title} {\bibinfo {title} {Elastic turbulence in a
  polymer solution flow},\ }\href@noop {} {\bibfield  {journal} {\bibinfo
  {journal} {Nature}\ }\textbf {\bibinfo {volume} {405}},\ \bibinfo {pages}
  {53} (\bibinfo {year} {2000})}\BibitemShut {NoStop}%
\bibitem [{\citenamefont {Larson}(2000)}]{larson2000turbulence}%
  \BibitemOpen
  \bibfield  {author} {\bibinfo {author} {\bibfnamefont {R.~G.}\ \bibnamefont
  {Larson}},\ }\bibfield  {title} {\bibinfo {title} {Turbulence without
  inertia},\ }\href@noop {} {\bibfield  {journal} {\bibinfo  {journal}
  {Nature}\ }\textbf {\bibinfo {volume} {405}},\ \bibinfo {pages} {27}
  (\bibinfo {year} {2000})}\BibitemShut {NoStop}%
\bibitem [{\citenamefont {Boffetta}\ \emph {et~al.}(2003)\citenamefont
  {Boffetta}, \citenamefont {Celani},\ and\ \citenamefont
  {Musacchio}}]{boffetta2003two}%
  \BibitemOpen
  \bibfield  {author} {\bibinfo {author} {\bibfnamefont {G.}~\bibnamefont
  {Boffetta}}, \bibinfo {author} {\bibfnamefont {A.}~\bibnamefont {Celani}},\
  and\ \bibinfo {author} {\bibfnamefont {S.}~\bibnamefont {Musacchio}},\
  }\bibfield  {title} {\bibinfo {title} {Two-dimensional turbulence of dilute
  polymer solutions},\ }\href@noop {} {\bibfield  {journal} {\bibinfo
  {journal} {Physical review letters}\ }\textbf {\bibinfo {volume} {91}},\
  \bibinfo {pages} {034501} (\bibinfo {year} {2003})}\BibitemShut {NoStop}%
\bibitem [{\citenamefont {Groisman}\ and\ \citenamefont
  {Steinberg}(2004)}]{groisman2004elastic}%
  \BibitemOpen
  \bibfield  {author} {\bibinfo {author} {\bibfnamefont {A.}~\bibnamefont
  {Groisman}}\ and\ \bibinfo {author} {\bibfnamefont {V.}~\bibnamefont
  {Steinberg}},\ }\bibfield  {title} {\bibinfo {title} {Elastic turbulence in
  curvilinear flows of polymer solutions},\ }\href@noop {} {\bibfield
  {journal} {\bibinfo  {journal} {New Journal of Physics}\ }\textbf {\bibinfo
  {volume} {6}},\ \bibinfo {pages} {29} (\bibinfo {year} {2004})}\BibitemShut
  {NoStop}%
\bibitem [{\citenamefont {Burghelea}\ \emph {et~al.}(2007)\citenamefont
  {Burghelea}, \citenamefont {Segre},\ and\ \citenamefont
  {Steinberg}}]{burghelea2007elastic}%
  \BibitemOpen
  \bibfield  {author} {\bibinfo {author} {\bibfnamefont {T.}~\bibnamefont
  {Burghelea}}, \bibinfo {author} {\bibfnamefont {E.}~\bibnamefont {Segre}},\
  and\ \bibinfo {author} {\bibfnamefont {V.}~\bibnamefont {Steinberg}},\
  }\bibfield  {title} {\bibinfo {title} {Elastic turbulence in von karman
  swirling flow between two disks},\ }\href@noop {} {\bibfield  {journal}
  {\bibinfo  {journal} {Physics of fluids}\ }\textbf {\bibinfo {volume} {19}}
  (\bibinfo {year} {2007})}\BibitemShut {NoStop}%
\bibitem [{\citenamefont {Steinberg}(2019)}]{steinberg2019scaling}%
  \BibitemOpen
  \bibfield  {author} {\bibinfo {author} {\bibfnamefont {V.}~\bibnamefont
  {Steinberg}},\ }\bibfield  {title} {\bibinfo {title} {Scaling relations in
  elastic turbulence},\ }\href@noop {} {\bibfield  {journal} {\bibinfo
  {journal} {Physical review letters}\ }\textbf {\bibinfo {volume} {123}},\
  \bibinfo {pages} {234501} (\bibinfo {year} {2019})}\BibitemShut {NoStop}%
\bibitem [{\citenamefont {Steinberg}(2021)}]{steinberg2021elastic}%
  \BibitemOpen
  \bibfield  {author} {\bibinfo {author} {\bibfnamefont {V.}~\bibnamefont
  {Steinberg}},\ }\bibfield  {title} {\bibinfo {title} {Elastic turbulence: an
  experimental view on inertialess random flow},\ }\href@noop {} {\bibfield
  {journal} {\bibinfo  {journal} {Annual Review of Fluid Mechanics}\ }\textbf
  {\bibinfo {volume} {53}},\ \bibinfo {pages} {27} (\bibinfo {year}
  {2021})}\BibitemShut {NoStop}%
\bibitem [{\citenamefont {Lellep}\ \emph {et~al.}(2024)\citenamefont {Lellep},
  \citenamefont {Linkmann},\ and\ \citenamefont {Morozov}}]{lellep2024purely}%
  \BibitemOpen
  \bibfield  {author} {\bibinfo {author} {\bibfnamefont {M.}~\bibnamefont
  {Lellep}}, \bibinfo {author} {\bibfnamefont {M.}~\bibnamefont {Linkmann}},\
  and\ \bibinfo {author} {\bibfnamefont {A.}~\bibnamefont {Morozov}},\
  }\bibfield  {title} {\bibinfo {title} {Purely elastic turbulence in
  pressure-driven channel flows},\ }\href@noop {} {\bibfield  {journal}
  {\bibinfo  {journal} {Proceedings of the National Academy of Sciences}\
  }\textbf {\bibinfo {volume} {121}},\ \bibinfo {pages} {e2318851121} (\bibinfo
  {year} {2024})}\BibitemShut {NoStop}%
\bibitem [{\citenamefont {Singh}\ \emph {et~al.}(2024)\citenamefont {Singh},
  \citenamefont {Perlekar}, \citenamefont {Mitra},\ and\ \citenamefont
  {Rosti}}]{singh2024intermittency}%
  \BibitemOpen
  \bibfield  {author} {\bibinfo {author} {\bibfnamefont {R.~K.}\ \bibnamefont
  {Singh}}, \bibinfo {author} {\bibfnamefont {P.}~\bibnamefont {Perlekar}},
  \bibinfo {author} {\bibfnamefont {D.}~\bibnamefont {Mitra}},\ and\ \bibinfo
  {author} {\bibfnamefont {M.~E.}\ \bibnamefont {Rosti}},\ }\bibfield  {title}
  {\bibinfo {title} {Intermittency in the not-so-smooth elastic turbulence},\
  }\href@noop {} {\bibfield  {journal} {\bibinfo  {journal} {Nature
  Communications}\ }\textbf {\bibinfo {volume} {15}},\ \bibinfo {pages} {4070}
  (\bibinfo {year} {2024})}\BibitemShut {NoStop}%
\bibitem [{\citenamefont {Rota}\ \emph {et~al.}(2024)\citenamefont {Rota},
  \citenamefont {Amor}, \citenamefont {Le~Clainche},\ and\ \citenamefont
  {Rosti}}]{rota2024unified}%
  \BibitemOpen
  \bibfield  {author} {\bibinfo {author} {\bibfnamefont {G.~F.}\ \bibnamefont
  {Rota}}, \bibinfo {author} {\bibfnamefont {C.}~\bibnamefont {Amor}}, \bibinfo
  {author} {\bibfnamefont {S.}~\bibnamefont {Le~Clainche}},\ and\ \bibinfo
  {author} {\bibfnamefont {M.~E.}\ \bibnamefont {Rosti}},\ }\bibfield  {title}
  {\bibinfo {title} {Unified view of elastic and elasto-inertial turbulence in
  channel flows at low and moderate reynolds numbers},\ }\href@noop {}
  {\bibfield  {journal} {\bibinfo  {journal} {Phys Rev Fluids}\ }\textbf
  {\bibinfo {volume} {9}},\ \bibinfo {pages} {L122602} (\bibinfo {year}
  {2024})}\BibitemShut {NoStop}%
\bibitem [{\citenamefont {Singh}\ and\ \citenamefont
  {Rosti}(2023)}]{singh2024interplay}%
  \BibitemOpen
  \bibfield  {author} {\bibinfo {author} {\bibfnamefont {R.~K.}\ \bibnamefont
  {Singh}}\ and\ \bibinfo {author} {\bibfnamefont {M.~E.}\ \bibnamefont
  {Rosti}},\ }\bibfield  {title} {\bibinfo {title} {The interplay of inertia
  and elasticity in polymeric flows},\ }\href@noop {} {\bibfield  {journal}
  {\bibinfo  {journal} {arXiv preprint arXiv:2309.14752}\ } (\bibinfo {year}
  {2023})}\BibitemShut {NoStop}%
\bibitem [{fn()}]{fn}%
  \BibitemOpen
  \href@noop {} {\bibinfo  {journal} {See the supplementary material for
  additional details; it includes the references \cite{singh2024intermittency,
  fattal2004constitutive, Yerasi_2024, rosti2023large, abdelgawad2023scaling,
  aswathy2024dynamics, fouxon2003spectra, nguyen2016small, casciola2007energy,
  rosti2025effect}}\ }\BibitemShut {NoStop}%
\bibitem [{\citenamefont {Frisch}(1995)}]{frisch1995turbulence}%
  \BibitemOpen
\bibfield  {journal} {  }\bibfield  {author} {\bibinfo {author} {\bibfnamefont
  {U.}~\bibnamefont {Frisch}},\ }\href@noop {} {\emph {\bibinfo {title}
  {Turbulence: the legacy of AN Kolmogorov}}}\ (\bibinfo  {publisher}
  {Cambridge university press},\ \bibinfo {year} {1995})\BibitemShut {NoStop}%
\bibitem [{\citenamefont {Davidson}(2015)}]{davidson2015turbulence}%
  \BibitemOpen
  \bibfield  {author} {\bibinfo {author} {\bibfnamefont {P.}~\bibnamefont
  {Davidson}},\ }\href@noop {} {\emph {\bibinfo {title} {Turbulence: an
  introduction for scientists and engineers}}}\ (\bibinfo  {publisher} {Oxford
  university press},\ \bibinfo {year} {2015})\BibitemShut {NoStop}%
\bibitem [{\citenamefont {Chen}\ \emph {et~al.}(1993)\citenamefont {Chen},
  \citenamefont {Doolen}, \citenamefont {Herring}, \citenamefont {Kraichnan},
  \citenamefont {Orszag},\ and\ \citenamefont {She}}]{chen1993far}%
  \BibitemOpen
  \bibfield  {author} {\bibinfo {author} {\bibfnamefont {S.}~\bibnamefont
  {Chen}}, \bibinfo {author} {\bibfnamefont {G.}~\bibnamefont {Doolen}},
  \bibinfo {author} {\bibfnamefont {J.~R.}\ \bibnamefont {Herring}}, \bibinfo
  {author} {\bibfnamefont {R.~H.}\ \bibnamefont {Kraichnan}}, \bibinfo {author}
  {\bibfnamefont {S.~A.}\ \bibnamefont {Orszag}},\ and\ \bibinfo {author}
  {\bibfnamefont {Z.~S.}\ \bibnamefont {She}},\ }\bibfield  {title} {\bibinfo
  {title} {Far-dissipation range of turbulence},\ }\href@noop {} {\bibfield
  {journal} {\bibinfo  {journal} {Physical review letters}\ }\textbf {\bibinfo
  {volume} {70}},\ \bibinfo {pages} {3051} (\bibinfo {year}
  {1993})}\BibitemShut {NoStop}%
\bibitem [{\citenamefont {Schumacher}(2007)}]{schumacher2007sub}%
  \BibitemOpen
  \bibfield  {author} {\bibinfo {author} {\bibfnamefont {J.}~\bibnamefont
  {Schumacher}},\ }\bibfield  {title} {\bibinfo {title} {Sub-kolmogorov-scale
  fluctuations in fluid turbulence},\ }\href@noop {} {\bibfield  {journal}
  {\bibinfo  {journal} {Europhysics Letters}\ }\textbf {\bibinfo {volume}
  {80}},\ \bibinfo {pages} {54001} (\bibinfo {year} {2007})}\BibitemShut
  {NoStop}%
\bibitem [{\citenamefont {Khurshid}\ \emph {et~al.}(2018)\citenamefont
  {Khurshid}, \citenamefont {Donzis},\ and\ \citenamefont
  {Sreenivasan}}]{khurshid2018energy}%
  \BibitemOpen
  \bibfield  {author} {\bibinfo {author} {\bibfnamefont {S.}~\bibnamefont
  {Khurshid}}, \bibinfo {author} {\bibfnamefont {D.~A.}\ \bibnamefont
  {Donzis}},\ and\ \bibinfo {author} {\bibfnamefont {K.}~\bibnamefont
  {Sreenivasan}},\ }\bibfield  {title} {\bibinfo {title} {Energy spectrum in
  the dissipation range},\ }\href@noop {} {\bibfield  {journal} {\bibinfo
  {journal} {Physical Review Fluids}\ }\textbf {\bibinfo {volume} {3}},\
  \bibinfo {pages} {082601} (\bibinfo {year} {2018})}\BibitemShut {NoStop}%
\bibitem [{\citenamefont {Gorbunova}\ \emph {et~al.}(2020)\citenamefont
  {Gorbunova}, \citenamefont {Balarac}, \citenamefont {Bourgoin}, \citenamefont
  {Canet}, \citenamefont {Mordant},\ and\ \citenamefont
  {Rossetto}}]{gorbunova2020analysis}%
  \BibitemOpen
  \bibfield  {author} {\bibinfo {author} {\bibfnamefont {A.}~\bibnamefont
  {Gorbunova}}, \bibinfo {author} {\bibfnamefont {G.}~\bibnamefont {Balarac}},
  \bibinfo {author} {\bibfnamefont {M.}~\bibnamefont {Bourgoin}}, \bibinfo
  {author} {\bibfnamefont {L.}~\bibnamefont {Canet}}, \bibinfo {author}
  {\bibfnamefont {N.}~\bibnamefont {Mordant}},\ and\ \bibinfo {author}
  {\bibfnamefont {V.}~\bibnamefont {Rossetto}},\ }\bibfield  {title} {\bibinfo
  {title} {Analysis of the dissipative range of the energy spectrum in grid
  turbulence and in direct numerical simulations},\ }\href@noop {} {\bibfield
  {journal} {\bibinfo  {journal} {Physical Review Fluids}\ }\textbf {\bibinfo
  {volume} {5}},\ \bibinfo {pages} {044604} (\bibinfo {year}
  {2020})}\BibitemShut {NoStop}%
\bibitem [{\citenamefont {Iyer}\ \emph {et~al.}(2020)\citenamefont {Iyer},
  \citenamefont {Sreenivasan},\ and\ \citenamefont {Yeung}}]{iyer2020scaling}%
  \BibitemOpen
  \bibfield  {author} {\bibinfo {author} {\bibfnamefont {K.~P.}\ \bibnamefont
  {Iyer}}, \bibinfo {author} {\bibfnamefont {K.~R.}\ \bibnamefont
  {Sreenivasan}},\ and\ \bibinfo {author} {\bibfnamefont {P.}~\bibnamefont
  {Yeung}},\ }\bibfield  {title} {\bibinfo {title} {Scaling exponents saturate
  in three-dimensional isotropic turbulence},\ }\href@noop {} {\bibfield
  {journal} {\bibinfo  {journal} {Physical Review Fluids}\ }\textbf {\bibinfo
  {volume} {5}},\ \bibinfo {pages} {054605} (\bibinfo {year}
  {2020})}\BibitemShut {NoStop}%
\bibitem [{Fn2()}]{Fn2}%
  \BibitemOpen
  \href@noop {} {\bibinfo  {journal} {Following \cite{rosti2023large}, we
  expect that the contribution of the polymeric flux in the inertial range
  would decrease for larger $De$ and $Re_{\lambda}$}\ }\BibitemShut {NoStop}%
\bibitem [{\citenamefont {Berti}\ \emph {et~al.}(2008)\citenamefont {Berti},
  \citenamefont {Bistagnino}, \citenamefont {Boffetta}, \citenamefont
  {Celani},\ and\ \citenamefont {Musacchio}}]{berti2008two}%
  \BibitemOpen
\bibfield  {journal} {  }\bibfield  {author} {\bibinfo {author} {\bibfnamefont
  {S.}~\bibnamefont {Berti}}, \bibinfo {author} {\bibfnamefont
  {A.}~\bibnamefont {Bistagnino}}, \bibinfo {author} {\bibfnamefont
  {G.}~\bibnamefont {Boffetta}}, \bibinfo {author} {\bibfnamefont
  {A.}~\bibnamefont {Celani}},\ and\ \bibinfo {author} {\bibfnamefont
  {S.}~\bibnamefont {Musacchio}},\ }\bibfield  {title} {\bibinfo {title}
  {Two-dimensional elastic turbulence},\ }\href@noop {} {\bibfield  {journal}
  {\bibinfo  {journal} {Physical Review E—Statistical, Nonlinear, and Soft
  Matter Physics}\ }\textbf {\bibinfo {volume} {77}},\ \bibinfo {pages}
  {055306} (\bibinfo {year} {2008})}\BibitemShut {NoStop}%
\bibitem [{\citenamefont {Fouxon}\ and\ \citenamefont
  {Lebedev}(2003)}]{fouxon2003spectra}%
  \BibitemOpen
  \bibfield  {author} {\bibinfo {author} {\bibfnamefont {A.}~\bibnamefont
  {Fouxon}}\ and\ \bibinfo {author} {\bibfnamefont {V.}~\bibnamefont
  {Lebedev}},\ }\bibfield  {title} {\bibinfo {title} {Spectra of turbulence in
  dilute polymer solutions},\ }\href@noop {} {\bibfield  {journal} {\bibinfo
  {journal} {Physics of Fluids}\ }\textbf {\bibinfo {volume} {15}},\ \bibinfo
  {pages} {2060} (\bibinfo {year} {2003})}\BibitemShut {NoStop}%
\bibitem [{\citenamefont {Sreenivasan}\ and\ \citenamefont
  {Antonia}(1997)}]{sreenivasan1997phenomenology}%
  \BibitemOpen
  \bibfield  {author} {\bibinfo {author} {\bibfnamefont {K.~R.}\ \bibnamefont
  {Sreenivasan}}\ and\ \bibinfo {author} {\bibfnamefont {R.~A.}\ \bibnamefont
  {Antonia}},\ }\bibfield  {title} {\bibinfo {title} {The phenomenology of
  small-scale turbulence},\ }\href@noop {} {\bibfield  {journal} {\bibinfo
  {journal} {Annual review of fluid mechanics}\ }\textbf {\bibinfo {volume}
  {29}},\ \bibinfo {pages} {435} (\bibinfo {year} {1997})}\BibitemShut
  {NoStop}%
\bibitem [{\citenamefont {Ishihara}\ \emph {et~al.}(2009)\citenamefont
  {Ishihara}, \citenamefont {Gotoh},\ and\ \citenamefont
  {Kaneda}}]{ishihara2009study}%
  \BibitemOpen
  \bibfield  {author} {\bibinfo {author} {\bibfnamefont {T.}~\bibnamefont
  {Ishihara}}, \bibinfo {author} {\bibfnamefont {T.}~\bibnamefont {Gotoh}},\
  and\ \bibinfo {author} {\bibfnamefont {Y.}~\bibnamefont {Kaneda}},\
  }\bibfield  {title} {\bibinfo {title} {Study of high--reynolds number
  isotropic turbulence by direct numerical simulation},\ }\href@noop {}
  {\bibfield  {journal} {\bibinfo  {journal} {Annual review of fluid
  mechanics}\ }\textbf {\bibinfo {volume} {41}},\ \bibinfo {pages} {165}
  (\bibinfo {year} {2009})}\BibitemShut {NoStop}%
\bibitem [{\citenamefont {Davidson}\ \emph {et~al.}(2012)\citenamefont
  {Davidson}, \citenamefont {Kaneda},\ and\ \citenamefont
  {Sreenivasan}}]{davidson2012ten}%
  \BibitemOpen
  \bibfield  {author} {\bibinfo {author} {\bibfnamefont {P.~A.}\ \bibnamefont
  {Davidson}}, \bibinfo {author} {\bibfnamefont {Y.}~\bibnamefont {Kaneda}},\
  and\ \bibinfo {author} {\bibfnamefont {K.~R.}\ \bibnamefont {Sreenivasan}},\
  }\href@noop {} {\emph {\bibinfo {title} {Ten chapters in turbulence}}}\
  (\bibinfo  {publisher} {Cambridge University Press},\ \bibinfo {year}
  {2012})\BibitemShut {NoStop}%
\bibitem [{\citenamefont {Schumacher}\ \emph {et~al.}(2007)\citenamefont
  {Schumacher}, \citenamefont {Sreenivasan},\ and\ \citenamefont
  {Yakhot}}]{schumacher2007asymptotic}%
  \BibitemOpen
  \bibfield  {author} {\bibinfo {author} {\bibfnamefont {J.}~\bibnamefont
  {Schumacher}}, \bibinfo {author} {\bibfnamefont {K.~R.}\ \bibnamefont
  {Sreenivasan}},\ and\ \bibinfo {author} {\bibfnamefont {V.}~\bibnamefont
  {Yakhot}},\ }\bibfield  {title} {\bibinfo {title} {Asymptotic exponents from
  low-reynolds-number flows},\ }\href@noop {} {\bibfield  {journal} {\bibinfo
  {journal} {New Journal of Physics}\ }\textbf {\bibinfo {volume} {9}},\
  \bibinfo {pages} {89} (\bibinfo {year} {2007})}\BibitemShut {NoStop}%
\bibitem [{\citenamefont {Nelkin}(1981)}]{nelkin1981dissipation}%
  \BibitemOpen
  \bibfield  {author} {\bibinfo {author} {\bibfnamefont {M.}~\bibnamefont
  {Nelkin}},\ }\bibfield  {title} {\bibinfo {title} {Do the dissipation
  fluctuations in high reynolds number turbulence define a universal
  exponent?},\ }\href@noop {} {\bibfield  {journal} {\bibinfo  {journal}
  {Physics of Fluids}\ }\textbf {\bibinfo {volume} {24}},\ \bibinfo {pages}
  {556} (\bibinfo {year} {1981})}\BibitemShut {NoStop}%
\bibitem [{\citenamefont {Cates}\ and\ \citenamefont
  {Deutsch}(1987)}]{cates1987spatial}%
  \BibitemOpen
  \bibfield  {author} {\bibinfo {author} {\bibfnamefont {M.}~\bibnamefont
  {Cates}}\ and\ \bibinfo {author} {\bibfnamefont {J.}~\bibnamefont
  {Deutsch}},\ }\bibfield  {title} {\bibinfo {title} {Spatial correlations in
  multifractals},\ }\href@noop {} {\bibfield  {journal} {\bibinfo  {journal}
  {Physical Review A}\ }\textbf {\bibinfo {volume} {35}},\ \bibinfo {pages}
  {4907} (\bibinfo {year} {1987})}\BibitemShut {NoStop}%
\bibitem [{\citenamefont {Sreenivasan}\ and\ \citenamefont
  {Kailasnath}(1993)}]{sreenivasan1993update}%
  \BibitemOpen
  \bibfield  {author} {\bibinfo {author} {\bibfnamefont {K.}~\bibnamefont
  {Sreenivasan}}\ and\ \bibinfo {author} {\bibfnamefont {P.}~\bibnamefont
  {Kailasnath}},\ }\bibfield  {title} {\bibinfo {title} {An update on the
  intermittency exponent in turbulence},\ }\href@noop {} {\bibfield  {journal}
  {\bibinfo  {journal} {Physics of Fluids A: Fluid Dynamics}\ }\textbf
  {\bibinfo {volume} {5}},\ \bibinfo {pages} {512} (\bibinfo {year}
  {1993})}\BibitemShut {NoStop}%
\bibitem [{\citenamefont {Meneveau}\ and\ \citenamefont
  {Sreenivasan}(1991)}]{meneveau1991multifractal}%
  \BibitemOpen
  \bibfield  {author} {\bibinfo {author} {\bibfnamefont {C.}~\bibnamefont
  {Meneveau}}\ and\ \bibinfo {author} {\bibfnamefont {K.}~\bibnamefont
  {Sreenivasan}},\ }\bibfield  {title} {\bibinfo {title} {The multifractal
  nature of turbulent energy dissipation},\ }\href@noop {} {\bibfield
  {journal} {\bibinfo  {journal} {Journal of Fluid Mechanics}\ }\textbf
  {\bibinfo {volume} {224}},\ \bibinfo {pages} {429} (\bibinfo {year}
  {1991})}\BibitemShut {NoStop}%
\bibitem [{\citenamefont {Buaria}\ and\ \citenamefont
  {Sreenivasan}(2022)}]{buaria2022intermittency}%
  \BibitemOpen
  \bibfield  {author} {\bibinfo {author} {\bibfnamefont {D.}~\bibnamefont
  {Buaria}}\ and\ \bibinfo {author} {\bibfnamefont {K.~R.}\ \bibnamefont
  {Sreenivasan}},\ }\bibfield  {title} {\bibinfo {title} {Intermittency of
  turbulent velocity and scalar fields using three-dimensional local
  averaging},\ }\href@noop {} {\bibfield  {journal} {\bibinfo  {journal}
  {Physical Review Fluids}\ }\textbf {\bibinfo {volume} {7}},\ \bibinfo {pages}
  {L072601} (\bibinfo {year} {2022})}\BibitemShut {NoStop}%
\bibitem [{\citenamefont {Schumacher}\ \emph {et~al.}(2014)\citenamefont
  {Schumacher}, \citenamefont {Scheel}, \citenamefont {Krasnov}, \citenamefont
  {Donzis}, \citenamefont {Yakhot},\ and\ \citenamefont
  {Sreenivasan}}]{schumacher2014small}%
  \BibitemOpen
  \bibfield  {author} {\bibinfo {author} {\bibfnamefont {J.}~\bibnamefont
  {Schumacher}}, \bibinfo {author} {\bibfnamefont {J.~D.}\ \bibnamefont
  {Scheel}}, \bibinfo {author} {\bibfnamefont {D.}~\bibnamefont {Krasnov}},
  \bibinfo {author} {\bibfnamefont {D.~A.}\ \bibnamefont {Donzis}}, \bibinfo
  {author} {\bibfnamefont {V.}~\bibnamefont {Yakhot}},\ and\ \bibinfo {author}
  {\bibfnamefont {K.~R.}\ \bibnamefont {Sreenivasan}},\ }\bibfield  {title}
  {\bibinfo {title} {Small-scale universality in fluid turbulence},\
  }\href@noop {} {\bibfield  {journal} {\bibinfo  {journal} {Proceedings of the
  National Academy of Sciences}\ }\textbf {\bibinfo {volume} {111}},\ \bibinfo
  {pages} {10961} (\bibinfo {year} {2014})}\BibitemShut {NoStop}%
\bibitem [{\citenamefont {Tabor}\ and\ \citenamefont
  {De~Gennes}(1986)}]{tabor1986cascade}%
  \BibitemOpen
  \bibfield  {author} {\bibinfo {author} {\bibfnamefont {M.}~\bibnamefont
  {Tabor}}\ and\ \bibinfo {author} {\bibfnamefont {P.}~\bibnamefont
  {De~Gennes}},\ }\bibfield  {title} {\bibinfo {title} {A cascade theory of
  drag reduction},\ }\href@noop {} {\bibfield  {journal} {\bibinfo  {journal}
  {Europhysics Letters}\ }\textbf {\bibinfo {volume} {2}},\ \bibinfo {pages}
  {519} (\bibinfo {year} {1986})}\BibitemShut {NoStop}%
\bibitem [{\citenamefont {De~Gennes}(1986)}]{de1986towards}%
  \BibitemOpen
  \bibfield  {author} {\bibinfo {author} {\bibfnamefont {P.}~\bibnamefont
  {De~Gennes}},\ }\bibfield  {title} {\bibinfo {title} {Towards a scaling
  theory of drag reduction},\ }\href@noop {} {\bibfield  {journal} {\bibinfo
  {journal} {Physica A: Statistical Mechanics and its Applications}\ }\textbf
  {\bibinfo {volume} {140}},\ \bibinfo {pages} {9} (\bibinfo {year}
  {1986})}\BibitemShut {NoStop}%
\bibitem [{\citenamefont {Sreenivasan}\ and\ \citenamefont
  {White}(2000)}]{sreenivasan2000onset}%
  \BibitemOpen
  \bibfield  {author} {\bibinfo {author} {\bibfnamefont {K.~R.}\ \bibnamefont
  {Sreenivasan}}\ and\ \bibinfo {author} {\bibfnamefont {C.~M.}\ \bibnamefont
  {White}},\ }\bibfield  {title} {\bibinfo {title} {The onset of drag reduction
  by dilute polymer additives, and the maximum drag reduction asymptote},\
  }\href@noop {} {\bibfield  {journal} {\bibinfo  {journal} {Journal of Fluid
  Mechanics}\ }\textbf {\bibinfo {volume} {409}},\ \bibinfo {pages} {149}
  (\bibinfo {year} {2000})}\BibitemShut {NoStop}%
\bibitem [{\citenamefont {Lumley}(1969)}]{lumley1969drag}%
  \BibitemOpen
  \bibfield  {author} {\bibinfo {author} {\bibfnamefont {J.~L.}\ \bibnamefont
  {Lumley}},\ }\bibfield  {title} {\bibinfo {title} {Drag reduction by
  additives},\ }\href@noop {} {\bibfield  {journal} {\bibinfo  {journal}
  {Annual review of fluid mechanics}\ }\textbf {\bibinfo {volume} {1}}
  (\bibinfo {year} {1969})}\BibitemShut {NoStop}%
\bibitem [{\citenamefont {Lumley}(1973)}]{lumley1973drag}%
  \BibitemOpen
  \bibfield  {author} {\bibinfo {author} {\bibfnamefont {J.~L.}\ \bibnamefont
  {Lumley}},\ }\bibfield  {title} {\bibinfo {title} {Drag reduction in
  turbulent flow by polymer additives},\ }\href@noop {} {\bibfield  {journal}
  {\bibinfo  {journal} {Journal of Polymer Science: Macromolecular Reviews}\
  }\textbf {\bibinfo {volume} {7}},\ \bibinfo {pages} {263} (\bibinfo {year}
  {1973})}\BibitemShut {NoStop}%
\bibitem [{\citenamefont {Procaccia}\ \emph {et~al.}(2008)\citenamefont
  {Procaccia}, \citenamefont {L’vov},\ and\ \citenamefont
  {Benzi}}]{procaccia2008colloquium}%
  \BibitemOpen
  \bibfield  {author} {\bibinfo {author} {\bibfnamefont {I.}~\bibnamefont
  {Procaccia}}, \bibinfo {author} {\bibfnamefont {V.~S.}\ \bibnamefont
  {L’vov}},\ and\ \bibinfo {author} {\bibfnamefont {R.}~\bibnamefont
  {Benzi}},\ }\bibfield  {title} {\bibinfo {title} {Colloquium: Theory of drag
  reduction by polymers in wall-bounded turbulence},\ }\href@noop {} {\bibfield
   {journal} {\bibinfo  {journal} {Reviews of Modern Physics}\ }\textbf
  {\bibinfo {volume} {80}},\ \bibinfo {pages} {225} (\bibinfo {year}
  {2008})}\BibitemShut {NoStop}%
\bibitem [{\citenamefont {Balkovsky}\ \emph {et~al.}(2001)\citenamefont
  {Balkovsky}, \citenamefont {Fouxon},\ and\ \citenamefont
  {Lebedev}}]{balkovsky2001turbulence}%
  \BibitemOpen
  \bibfield  {author} {\bibinfo {author} {\bibfnamefont {E.}~\bibnamefont
  {Balkovsky}}, \bibinfo {author} {\bibfnamefont {A.}~\bibnamefont {Fouxon}},\
  and\ \bibinfo {author} {\bibfnamefont {V.}~\bibnamefont {Lebedev}},\
  }\bibfield  {title} {\bibinfo {title} {Turbulence of polymer solutions},\
  }\href@noop {} {\bibfield  {journal} {\bibinfo  {journal} {Physical Review
  E}\ }\textbf {\bibinfo {volume} {64}},\ \bibinfo {pages} {056301} (\bibinfo
  {year} {2001})}\BibitemShut {NoStop}%
\bibitem [{\citenamefont {Datta}\ \emph {et~al.}(2022)\citenamefont {Datta},
  \citenamefont {Ardekani}, \citenamefont {Arratia}, \citenamefont {Beris},
  \citenamefont {Bischofberger}, \citenamefont {McKinley}, \citenamefont
  {Eggers}, \citenamefont {L{\'o}pez-Aguilar}, \citenamefont {Fielding},
  \citenamefont {Frishman} \emph {et~al.}}]{datta2022perspectives}%
  \BibitemOpen
  \bibfield  {author} {\bibinfo {author} {\bibfnamefont {S.~S.}\ \bibnamefont
  {Datta}}, \bibinfo {author} {\bibfnamefont {A.~M.}\ \bibnamefont {Ardekani}},
  \bibinfo {author} {\bibfnamefont {P.~E.}\ \bibnamefont {Arratia}}, \bibinfo
  {author} {\bibfnamefont {A.~N.}\ \bibnamefont {Beris}}, \bibinfo {author}
  {\bibfnamefont {I.}~\bibnamefont {Bischofberger}}, \bibinfo {author}
  {\bibfnamefont {G.~H.}\ \bibnamefont {McKinley}}, \bibinfo {author}
  {\bibfnamefont {J.~G.}\ \bibnamefont {Eggers}}, \bibinfo {author}
  {\bibfnamefont {J.~E.}\ \bibnamefont {L{\'o}pez-Aguilar}}, \bibinfo {author}
  {\bibfnamefont {S.~M.}\ \bibnamefont {Fielding}}, \bibinfo {author}
  {\bibfnamefont {A.}~\bibnamefont {Frishman}}, \emph {et~al.},\ }\bibfield
  {title} {\bibinfo {title} {Perspectives on viscoelastic flow instabilities
  and elastic turbulence},\ }\href@noop {} {\bibfield  {journal} {\bibinfo
  {journal} {Physical Review Fluids}\ }\textbf {\bibinfo {volume} {7}},\
  \bibinfo {pages} {080701} (\bibinfo {year} {2022})}\BibitemShut {NoStop}%
\bibitem [{\citenamefont {Haward}\ \emph {et~al.}(2021)\citenamefont {Haward},
  \citenamefont {Hopkins},\ and\ \citenamefont {Shen}}]{haward2021stagnation}%
  \BibitemOpen
  \bibfield  {author} {\bibinfo {author} {\bibfnamefont {S.~J.}\ \bibnamefont
  {Haward}}, \bibinfo {author} {\bibfnamefont {C.~C.}\ \bibnamefont
  {Hopkins}},\ and\ \bibinfo {author} {\bibfnamefont {A.~Q.}\ \bibnamefont
  {Shen}},\ }\bibfield  {title} {\bibinfo {title} {Stagnation points control
  chaotic fluctuations in viscoelastic porous media flow},\ }\href@noop {}
  {\bibfield  {journal} {\bibinfo  {journal} {Proceedings of the National
  Academy of Sciences}\ }\textbf {\bibinfo {volume} {118}},\ \bibinfo {pages}
  {e2111651118} (\bibinfo {year} {2021})}\BibitemShut {NoStop}%
\bibitem [{\citenamefont {Browne}\ and\ \citenamefont
  {Datta}(2024)}]{browne2024harnessing}%
  \BibitemOpen
  \bibfield  {author} {\bibinfo {author} {\bibfnamefont {C.~A.}\ \bibnamefont
  {Browne}}\ and\ \bibinfo {author} {\bibfnamefont {S.~S.}\ \bibnamefont
  {Datta}},\ }\bibfield  {title} {\bibinfo {title} {Harnessing elastic
  instabilities for enhanced mixing and reaction kinetics in porous media},\
  }\href@noop {} {\bibfield  {journal} {\bibinfo  {journal} {Proceedings of the
  National Academy of Sciences}\ }\textbf {\bibinfo {volume} {121}},\ \bibinfo
  {pages} {e2320962121} (\bibinfo {year} {2024})}\BibitemShut {NoStop}%
\bibitem [{\citenamefont {Fattal}\ and\ \citenamefont
  {Kupferman}(2004)}]{fattal2004constitutive}%
  \BibitemOpen
  \bibfield  {author} {\bibinfo {author} {\bibfnamefont {R.}~\bibnamefont
  {Fattal}}\ and\ \bibinfo {author} {\bibfnamefont {R.}~\bibnamefont
  {Kupferman}},\ }\bibfield  {title} {\bibinfo {title} {Constitutive laws for
  the matrix-logarithm of the conformation tensor},\ }\href@noop {} {\bibfield
  {journal} {\bibinfo  {journal} {Journal of Non-Newtonian Fluid Mechanics}\
  }\textbf {\bibinfo {volume} {123}},\ \bibinfo {pages} {281} (\bibinfo {year}
  {2004})}\BibitemShut {NoStop}%
\bibitem [{\citenamefont {Yerasi}\ \emph {et~al.}(2024)\citenamefont {Yerasi},
  \citenamefont {Picardo}, \citenamefont {Gupta},\ and\ \citenamefont
  {Vincenzi}}]{Yerasi_2024}%
  \BibitemOpen
  \bibfield  {author} {\bibinfo {author} {\bibfnamefont {S.~R.}\ \bibnamefont
  {Yerasi}}, \bibinfo {author} {\bibfnamefont {J.~R.}\ \bibnamefont {Picardo}},
  \bibinfo {author} {\bibfnamefont {A.}~\bibnamefont {Gupta}},\ and\ \bibinfo
  {author} {\bibfnamefont {D.}~\bibnamefont {Vincenzi}},\ }\bibfield  {title}
  {\bibinfo {title} {Preserving large-scale features in simulations of elastic
  turbulence},\ }\href@noop {} {\bibfield  {journal} {\bibinfo  {journal}
  {Journal of Fluid Mechanics}\ }\textbf {\bibinfo {volume} {1000}},\ \bibinfo
  {pages} {A37} (\bibinfo {year} {2024})}\BibitemShut {NoStop}%
\bibitem [{\citenamefont {Abdelgawad}\ \emph {et~al.}(2023)\citenamefont
  {Abdelgawad}, \citenamefont {Cannon},\ and\ \citenamefont
  {Rosti}}]{abdelgawad2023scaling}%
  \BibitemOpen
  \bibfield  {author} {\bibinfo {author} {\bibfnamefont {M.~S.}\ \bibnamefont
  {Abdelgawad}}, \bibinfo {author} {\bibfnamefont {I.}~\bibnamefont {Cannon}},\
  and\ \bibinfo {author} {\bibfnamefont {M.~E.}\ \bibnamefont {Rosti}},\
  }\bibfield  {title} {\bibinfo {title} {Scaling and intermittency in turbulent
  flows of elastoviscoplastic fluids},\ }\href@noop {} {\bibfield  {journal}
  {\bibinfo  {journal} {Nature Physics}\ }\textbf {\bibinfo {volume} {19}},\
  \bibinfo {pages} {1059} (\bibinfo {year} {2023})}\BibitemShut {NoStop}%
\bibitem [{\citenamefont {Aswathy}\ and\ \citenamefont
  {Rosti}(2024)}]{aswathy2024dynamics}%
  \BibitemOpen
  \bibfield  {author} {\bibinfo {author} {\bibfnamefont {M.}~\bibnamefont
  {Aswathy}}\ and\ \bibinfo {author} {\bibfnamefont {M.~E.}\ \bibnamefont
  {Rosti}},\ }\bibfield  {title} {\bibinfo {title} {The dynamics of fibres
  dispersed in viscoelastic turbulent flows},\ }\href@noop {} {\bibfield
  {journal} {\bibinfo  {journal} {Journal of Fluid Mechanics}\ }\textbf
  {\bibinfo {volume} {984}},\ \bibinfo {pages} {A72} (\bibinfo {year}
  {2024})}\BibitemShut {NoStop}%
\bibitem [{\citenamefont {Casciola}\ and\ \citenamefont
  {De~Angelis}(2007)}]{casciola2007energy}%
  \BibitemOpen
  \bibfield  {author} {\bibinfo {author} {\bibfnamefont {C.~M.}\ \bibnamefont
  {Casciola}}\ and\ \bibinfo {author} {\bibfnamefont {E.}~\bibnamefont
  {De~Angelis}},\ }\bibfield  {title} {\bibinfo {title} {Energy transfer in
  turbulent polymer solutions},\ }\href@noop {} {\bibfield  {journal} {\bibinfo
   {journal} {Journal of Fluid Mechanics}\ }\textbf {\bibinfo {volume} {581}},\
  \bibinfo {pages} {419} (\bibinfo {year} {2007})}\BibitemShut {NoStop}%
\bibitem [{\citenamefont {Rosti}(2025)}]{rosti2025effect}%
  \BibitemOpen
  \bibfield  {author} {\bibinfo {author} {\bibfnamefont {M.~E.}\ \bibnamefont
  {Rosti}},\ }\bibfield  {title} {\bibinfo {title} {The effect of
  shear-thinning on the scalings and small-scale structures of turbulence},\
  }\href@noop {} {\bibfield  {journal} {\bibinfo  {journal} {Journal of Fluid
  Mechanics}\ }\textbf {\bibinfo {volume} {1012}},\ \bibinfo {pages} {R5}
  (\bibinfo {year} {2025})}\BibitemShut {NoStop}%
\end{thebibliography}%


\providecommand{\noopsort}[1]{}\providecommand{\singleletter}[1]{#1}%
\begin{thebibliography}{11}
\expandafter\ifx\csname natexlab\endcsname\relax\def\natexlab#1{#1}\fi
\expandafter\ifx\csname bibnamefont\endcsname\relax
  \def\bibnamefont#1{#1}\fi
\expandafter\ifx\csname bibfnamefont\endcsname\relax
  \def\bibfnamefont#1{#1}\fi
\expandafter\ifx\csname citenamefont\endcsname\relax
  \def\citenamefont#1{#1}\fi
\expandafter\ifx\csname url\endcsname\relax
  \def\url#1{\texttt{#1}}\fi
\expandafter\ifx\csname urlprefix\endcsname\relax\def\urlprefix{URL }\fi
\providecommand{\bibinfo}[2]{#2}
\providecommand{\eprint}[2][]{\url{#2}}

\bibitem[{\citenamefont{Singh et~al.}(2024)\citenamefont{Singh, Perlekar,
  Mitra, and Rosti}}]{singh2024intermittency}
\bibinfo{author}{\bibfnamefont{R.~K.} \bibnamefont{Singh}},
  \bibinfo{author}{\bibfnamefont{P.}~\bibnamefont{Perlekar}},
  \bibinfo{author}{\bibfnamefont{D.}~\bibnamefont{Mitra}}, \bibnamefont{and}
  \bibinfo{author}{\bibfnamefont{M.~E.} \bibnamefont{Rosti}},
  \bibinfo{journal}{Nature Communications} \textbf{\bibinfo{volume}{15}},
  \bibinfo{pages}{4070} (\bibinfo{year}{2024}).

\bibitem[{\citenamefont{Fattal and Kupferman}(2004)}]{fattal2004constitutive}
\bibinfo{author}{\bibfnamefont{R.}~\bibnamefont{Fattal}} \bibnamefont{and}
  \bibinfo{author}{\bibfnamefont{R.}~\bibnamefont{Kupferman}},
  \bibinfo{journal}{Journal of Non-Newtonian Fluid Mechanics}
  \textbf{\bibinfo{volume}{123}}, \bibinfo{pages}{281} (\bibinfo{year}{2004}).

\bibitem[{\citenamefont{Yerasi et~al.}(2024)\citenamefont{Yerasi, Picardo,
  Gupta, and Vincenzi}}]{Yerasi_2024}
\bibinfo{author}{\bibfnamefont{S.~R.} \bibnamefont{Yerasi}},
  \bibinfo{author}{\bibfnamefont{J.~R.} \bibnamefont{Picardo}},
  \bibinfo{author}{\bibfnamefont{A.}~\bibnamefont{Gupta}}, \bibnamefont{and}
  \bibinfo{author}{\bibfnamefont{D.}~\bibnamefont{Vincenzi}},
  \bibinfo{journal}{Journal of Fluid Mechanics}
  \textbf{\bibinfo{volume}{1000}}, \bibinfo{pages}{A37} (\bibinfo{year}{2024}).

\bibitem[{\citenamefont{Rosti et~al.}(2023)\citenamefont{Rosti, Perlekar, and
  Mitra}}]{rosti2023large}
\bibinfo{author}{\bibfnamefont{M.~E.} \bibnamefont{Rosti}},
  \bibinfo{author}{\bibfnamefont{P.}~\bibnamefont{Perlekar}}, \bibnamefont{and}
  \bibinfo{author}{\bibfnamefont{D.}~\bibnamefont{Mitra}},
  \bibinfo{journal}{Science Advances} \textbf{\bibinfo{volume}{9}},
  \bibinfo{pages}{eadd3831} (\bibinfo{year}{2023}).

\bibitem[{\citenamefont{Abdelgawad et~al.}(2023)\citenamefont{Abdelgawad,
  Cannon, and Rosti}}]{abdelgawad2023scaling}
\bibinfo{author}{\bibfnamefont{M.~S.} \bibnamefont{Abdelgawad}},
  \bibinfo{author}{\bibfnamefont{I.}~\bibnamefont{Cannon}}, \bibnamefont{and}
  \bibinfo{author}{\bibfnamefont{M.~E.} \bibnamefont{Rosti}},
  \bibinfo{journal}{Nature Physics} \textbf{\bibinfo{volume}{19}},
  \bibinfo{pages}{1059} (\bibinfo{year}{2023}).

\bibitem[{\citenamefont{Singh and Rosti}(2023)}]{singh2024interplay}
\bibinfo{author}{\bibfnamefont{R.~K.} \bibnamefont{Singh}} \bibnamefont{and}
  \bibinfo{author}{\bibfnamefont{M.~E.} \bibnamefont{Rosti}},
  \bibinfo{journal}{arXiv preprint arXiv:2309.14752}  (\bibinfo{year}{2023}).

\bibitem[{\citenamefont{Aswathy and Rosti}(2024)}]{aswathy2024dynamics}
\bibinfo{author}{\bibfnamefont{M.}~\bibnamefont{Aswathy}} \bibnamefont{and}
  \bibinfo{author}{\bibfnamefont{M.~E.} \bibnamefont{Rosti}},
  \bibinfo{journal}{Journal of Fluid Mechanics} \textbf{\bibinfo{volume}{984}},
  \bibinfo{pages}{A72} (\bibinfo{year}{2024}).

\bibitem[{\citenamefont{Nguyen et~al.}(2016)\citenamefont{Nguyen, Delache,
  Simo{\"e}ns, Bos, and El~Hajem}}]{nguyen2016small}
\bibinfo{author}{\bibfnamefont{M.~Q.} \bibnamefont{Nguyen}},
  \bibinfo{author}{\bibfnamefont{A.}~\bibnamefont{Delache}},
  \bibinfo{author}{\bibfnamefont{S.}~\bibnamefont{Simo{\"e}ns}},
  \bibinfo{author}{\bibfnamefont{W.~J.} \bibnamefont{Bos}}, \bibnamefont{and}
  \bibinfo{author}{\bibfnamefont{M.}~\bibnamefont{El~Hajem}},
  \bibinfo{journal}{Physical Review Fluids} \textbf{\bibinfo{volume}{1}},
  \bibinfo{pages}{083301} (\bibinfo{year}{2016}).

\bibitem[{\citenamefont{Rosti}(2025)}]{rosti2025effect}
\bibinfo{author}{\bibfnamefont{M.~E.} \bibnamefont{Rosti}},
  \bibinfo{journal}{Journal of Fluid Mechanics}
  \textbf{\bibinfo{volume}{1012}}, \bibinfo{pages}{R5} (\bibinfo{year}{2025}).

\bibitem[{\citenamefont{Casciola and De~Angelis}(2007)}]{casciola2007energy}
\bibinfo{author}{\bibfnamefont{C.~M.} \bibnamefont{Casciola}} \bibnamefont{and}
  \bibinfo{author}{\bibfnamefont{E.}~\bibnamefont{De~Angelis}},
  \bibinfo{journal}{Journal of Fluid Mechanics} \textbf{\bibinfo{volume}{581}},
  \bibinfo{pages}{419} (\bibinfo{year}{2007}).

\bibitem[{\citenamefont{Fouxon and Lebedev}(2003)}]{fouxon2003spectra}
\bibinfo{author}{\bibfnamefont{A.}~\bibnamefont{Fouxon}} \bibnamefont{and}
  \bibinfo{author}{\bibfnamefont{V.}~\bibnamefont{Lebedev}},
  \bibinfo{journal}{Physics of Fluids} \textbf{\bibinfo{volume}{15}},
  \bibinfo{pages}{2060} (\bibinfo{year}{2003}).

\end{thebibliography}
\end{document}



\setcounter{table}{0}
\makeatletter 
\renewcommand{\thetable}{S\@arabic\c@table}
\makeatother

\setcounter{figure}{0}
\makeatletter 
\renewcommand{\thefigure}{S\@arabic\c@figure}
\makeatother

\setcounter{equation}{0}
\makeatletter 
\renewcommand{\theequation}{S\@arabic\c@equation}
\makeatother


\title{Supplementary Material for ``Elastic turbulence hides in the small scales of inertial polymeric turbulence"}


\affiliation{}




\maketitle

\section{Details of the numerical simulation}

\begin{figure}
		\includegraphics[scale=0.6]{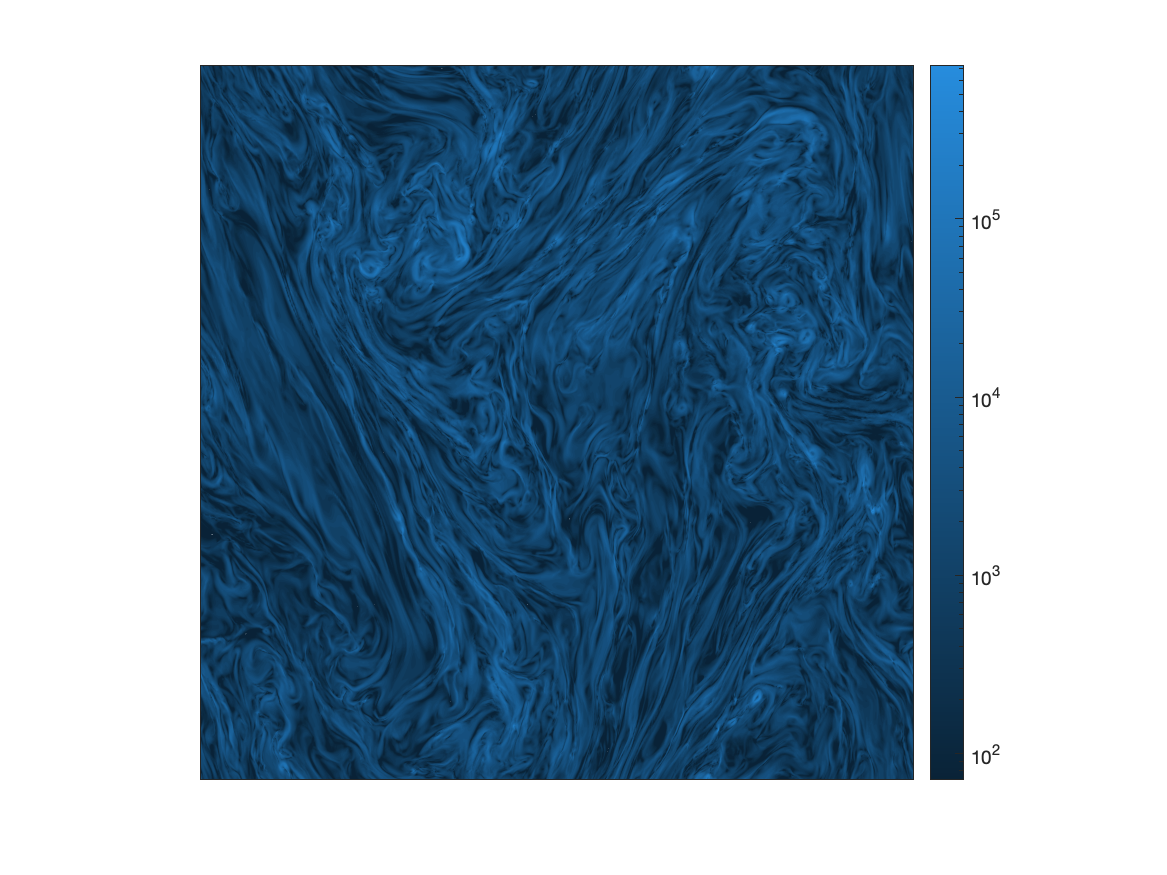}
	\caption{An instantaneous 2D snapshot of the magnitude of the vorticity field in polymeric turbulence at $Re_\lambda \approx 240$ and $De \approx 9$.}
    \label{fig:snapshot}
\end{figure}

The governing equations are solved in a periodic domain of length $2 \pi $, and the turbulent state is maintained using the ABC forcing with $F = \mu  ( (\sin z + \cos y) \hat{\boldsymbol{x}} + (\sin x + \cos z) \hat{\boldsymbol{y}} + (\sin y + \cos x) \hat{\boldsymbol{z}})$ \cite{singh2024intermittency}. We use the in house flow solver \textit{Fujin}, where a second order finite difference scheme is used for the spatial derivatives, while the Adams-Bashforth scheme is used for the time integration. We use the log-conformation approach to handle the stress constitutive equation, which has been shown to be reliable at large $De$ and hence we do not add any artificial stress diffusion, and avoid any spurious artifacts the latter can introduce \cite{fattal2004constitutive, Yerasi_2024}. The chosen methodology has been extensively validated \cite{rosti2023large, abdelgawad2023scaling, singh2024intermittency, singh2024interplay, aswathy2024dynamics}. For the polymeric turbulence case ($Re_\lambda \approx 240$, $De \approx 9$), we have used the unprecedentedly high resolution of $N=2048^3$ to ensure that the largest wavenumeber resolved lies sufficiently in the dissipation range, i.e., we have more than one decade of dissipation range ($k_{max} \eta \sim 13.4$). This is needed to do statistical analysis and obtain intermittency exponents below the Kolmogorov scale (i.e in the range $\eta^{-1} <k <k_{max}$). For all the other simulations we use the lower resolution $N=1024^3$, since detailed resolution of the dissipation range is not required in these cases. Note that for the elastic turbulence case, we choose $Re_{\lambda} \sim 40$ to be consistent with recent results \cite{singh2024intermittency}, where it was shown that inertial effects are negligible at this $Re_{\lambda}$. All the results presented here are averaged over multiple snapshots of the fields, obtained after a statistically stationary state has been reached for long times; specifically, for the polymeric turbulence case ($Re_\lambda \approx 240, De \approx 9$) 7 snapshots were used to obtain the averages, spanning one eddy turnover time at the largest scale. We have verified that one eddy turnover time is sufficient by calculating the statistics over multiple eddy turnover times at the lower resolution ($1024^3$). Note that, we have defined the Taylor Reynolds number $Re_{\lambda}$ (and the Taylor length scale) for all the cases with the fluid dissipation, given by $\epsilon_f = 2 \nu_f  \boldsymbol{S} \boldsymbol{:} \boldsymbol{S}$, with $\boldsymbol{S} = 1/2 (\nabla \boldsymbol{u}+ (\nabla \boldsymbol{u})^{T} )$. Finally, we have used the same ratio of the fluid viscosity to the total solution viscosity across all the parameters, $\beta = \frac{\mu_f}{(\mu_f+\mu_p)} = 0.9$; which is appropriate for dilute polymer solutions. Figure \ref{fig:snapshot} shows an instantaneous 2D snapshot of the magnitude of the vorticity field obtained using the simulation. 

\section{Derivation of the energy flux equation}
The relevant energy flux equation has been derived recently in \cite{rosti2023large, singh2024interplay}. Hence, here we only give a brief overview. The equation for the energy flux can be derived by first Fourier transforming the momentum equation to obtain,
\begin{equation}
\rho_f (\partial_t \boldsymbol{\hat{u}} + \boldsymbol{\hat{N}}) = - i \boldsymbol{k} \hat{p} + \mu_p/\tau_p i\boldsymbol{k} \boldsymbol{.} \boldsymbol{\hat{C}} -  \mu_f k^{2} \boldsymbol{\hat{u}} + \boldsymbol{F} ,
\label{eq:flux1}
\end{equation}
where $\boldsymbol{\hat{N}}$ is the Fourier transform of the non-linear convective term. Next, we take the complex conjugate of \eqref{eq:flux1},
\begin{equation}
\rho_f (\partial_t \boldsymbol{\hat{u}}^{*} + \boldsymbol{\hat{N}}^{*}) = i \boldsymbol{k} \hat{p}^{*} - \mu_p/\tau_p i \boldsymbol{k} \boldsymbol{.} \boldsymbol{\hat{C}}^{*} -  \mu_f k^{2} \boldsymbol{\hat{u}}^{*} + \boldsymbol{F}^{*} .
\label{eq:flux2}
\end{equation}
Recall that the turbulent kinetic energy is $\hat{E} =(1/2) \rho_f \boldsymbol{\hat{u}} \boldsymbol{.} \boldsymbol{\hat{u}}^{*} $. Hence, the final equation is given by $1/2$($\boldsymbol{\hat{u}}^{*} \boldsymbol{.} $ \eqref{eq:flux1} + $\boldsymbol{\hat{u}} \boldsymbol{.}$ \eqref{eq:flux2}), which after simplification leads to 
\begin{equation}
\partial_t \hat{E} = \hat{T} + \hat{P} + \hat{V}  + \hat{F_{e}} 
\end{equation}
where $\hat{T}$ is the non linear inertial term given by $ \hat{T} = -1/2 \rho_f (\boldsymbol{\hat{u}}^{*} \boldsymbol{.} \boldsymbol{\hat{N}} + \boldsymbol{\hat{u}} \boldsymbol{.} \boldsymbol{\hat{N}}^{*})$, $\hat{P}$ is the polymeric term given by $ \hat{P} = 1/2 \mu_f/\tau_p i (\boldsymbol{\hat{u}}^{*} \boldsymbol{.} (\boldsymbol{k} \boldsymbol{.} \boldsymbol{\hat{C}}) - \boldsymbol{\hat{u}} \boldsymbol{.} (\boldsymbol{k} \boldsymbol{.} \boldsymbol{\hat{C}}^{*}))$,  $\hat{V}$ is the viscous term given by $\hat{V} = - 2 \mu_f k^2 \hat{E}  $ and finally $\hat{F_{e}}$ is due to the external force given by $\hat{F_{e}} = 1/2 (\boldsymbol{\hat{u}}^{*} \boldsymbol{.} \boldsymbol{\hat{F}} + \boldsymbol{\hat{u}} \boldsymbol{.} \boldsymbol{\hat{F}}^{*})$. Integrating over the angular degrees of freedom and from $0$ to $k$, and assuming that the flow achieves a statistically stationary for long times, we arrive at the energy flux equation at the scale $k$ as:
\begin{equation}
    0 = \Pi_f + \mathcal{P} +\mathcal{D}_f + \mathcal{F}_e
\end{equation}
where $\Pi_f = \int_0^k \mathrm{d} k \mathrm{d}\Omega \hat{T}$ is the flux due to the non-linear inertial term, $\mathcal{P} = \int_0^k \mathrm{d} k \mathrm{d}\Omega \hat{P}$ is the energy flux due to polymers, $\mathcal{D}_f = \int_0^k \mathrm{d} k \mathrm{d}\Omega \hat{V}$ is the dissipative flux due to viscosity and finally $\mathcal{F}_e = \int_0^k \mathrm{d} k \mathrm{d}\Omega \hat{F}_e$ is the flux due to the external force. Note that $\mathcal{F}_e$ only contributes at the largest wavenumbers. 

Following \cite{rosti2023large}, the polymeric flux $\mathcal{P}$ can further be divided into a dissipative ($\mathcal{D}_p$) and non-dissipative ($\Pi_p$) component by defining,
\begin{equation}
    \mathcal{D}_p = \frac{\epsilon_p}{\epsilon_f} \mathcal{D}_f
\end{equation}
such that for $k \rightarrow \infty$, using $\mathcal{D}_f \rightarrow \epsilon_f$, we have $\mathcal{D}_p \rightarrow \mathcal{P} \rightarrow \epsilon_p$.

\section{Comparison with FENE-P}
\begin{figure}[hbt!]
		\includegraphics[scale=0.55]{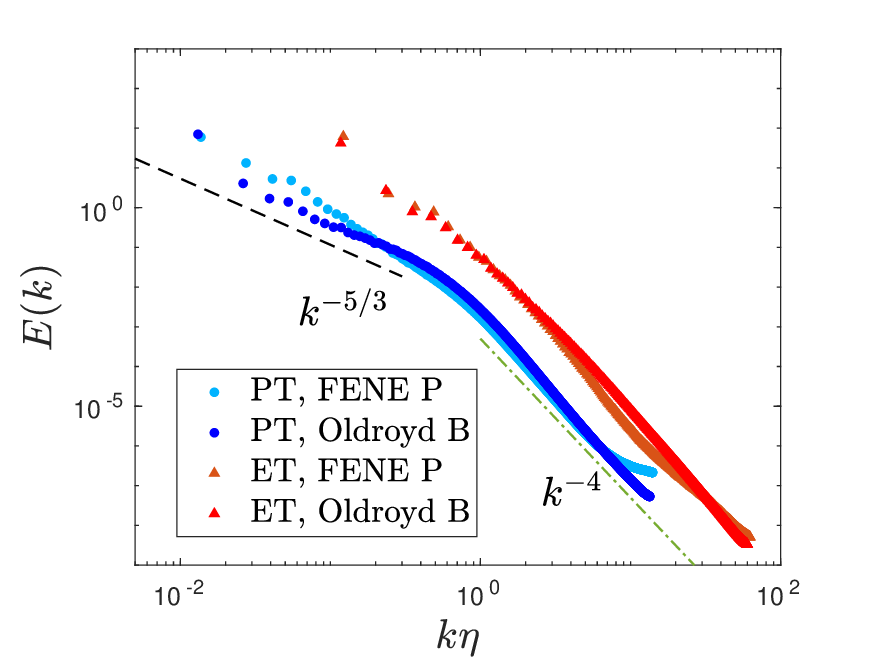}
	\caption{The comparison of the energy spectrum $E(k)$ between the Oldroyd B and FENE-P constitutive models for polymeric turbulence ($Re_\lambda \approx 240 $, $De \approx 9 $) and elastic turbulence ($Re_\lambda \approx 40 $, $De \approx 9 $). Here, the extensibility parameter and the viscosity ratio are $L= 60$ and $\beta = 0.9$, respectively.}
    \label{supp:fene}
\end{figure}
To confirm that the results presented in the manuscript are independent of the constitutive model, here we present a comparison with the results obtained using the FENE-P model. Existing results already indicate that Oldroyd-B captures all the essential physics in both PT and ET \cite{rosti2023large, singh2024intermittency, Yerasi_2024, nguyen2016small}. For both the PT and ET FENE-P cases, we fix the extensibility parameter to $L = 60$, and the viscosity ratio to $\beta = 0.9$. For PT FENE-P, we again consider the large resolution with $N=2048^3$  grid points, while for ET FENE-P we use $1024^3$. Figure \ref{supp:fene} shows the energy spectrum $E(k)$ versus the non dimensional wavenumber $k\eta$ for Oldroyd-B versus FENE-P. For PT, we essentially see the same $k^{-4}$ scaling for $k\eta \gg 1$ in both the cases, thus confirming model independence of the results presented. Note that the FENE-P case is seen to require greater resolution than Oldroyd-B to resolve the smallest scales of the flow, due to the shear-thinning \cite{rosti2025effect}.

\section{Energy flux for NT}
\begin{figure}[hbt!]
		\includegraphics[scale=0.55]{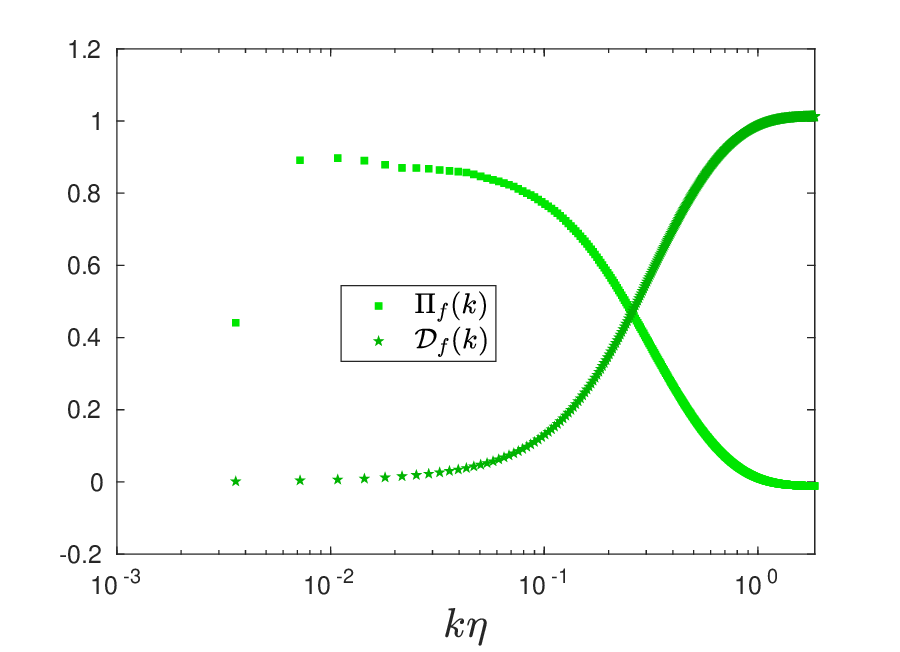}
	\caption{The various energy fluxes in Fourier space versus $k \eta$ for Newtonian turbulence ($NT$).}
    \label{supp:flux_nt}
\end{figure}

Figure \ref{supp:flux_nt} shows the variation of the inertial ($\Pi_f$) and viscous ($D_f$) fluxes with the non-dimensional wavenumber ($k\eta$) for the case of Newtonian turbulence. 

\section{Polymeric Energy Spectrum ($E_p(k)$)}
\begin{figure}[hbt!]
		\includegraphics[scale=0.55]{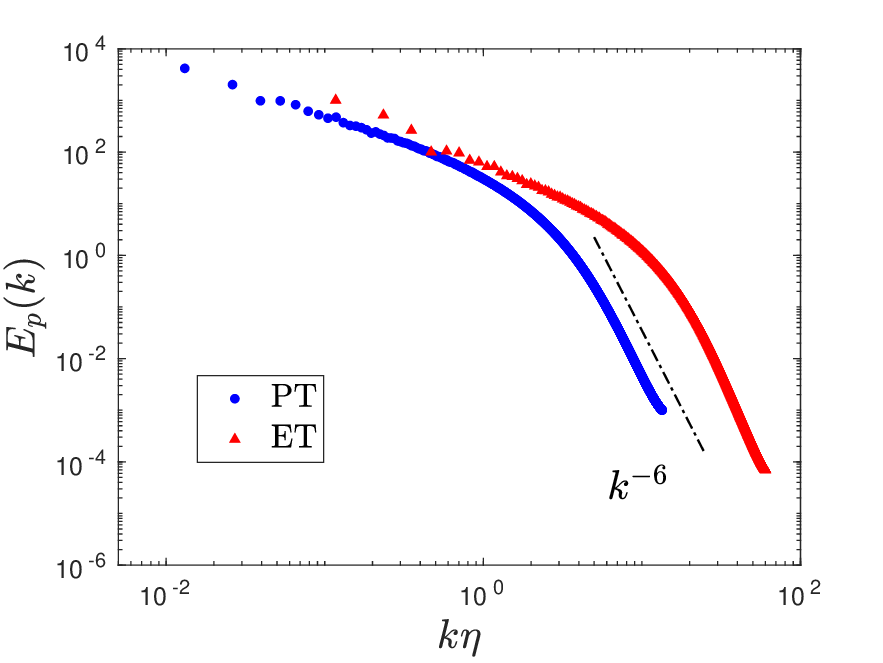}
	\caption{Polymeric energy spectrum $E_p(k)$ versus the non-dimensional wavenumber $k \eta$, for polymeric (PT) and elastic (ET) turbulence (NT).}
    \label{supp:spec_pol}
\end{figure}
In addition to the fluid kinetic energy, the spectrum of the energy stored in the polymers is given by
\begin{equation}
    E_p(k) =  \frac{\mu_p}{2 \tau_p}  \int_{\vert \boldsymbol{k} \vert = k}B_{ij}(\boldsymbol{k}) B_{ji}(\boldsymbol{-k})\mathrm{d}\Omega_{k},
\end{equation}
where $B_{ij}$ is the matrix square root matrix of $C_{ij}$, and $\Omega_k$ denotes the angular integral \cite{casciola2007energy, nguyen2016small, singh2024interplay}. Figure \ref{supp:spec_pol} shows the polymeric energy spectra $E_p(k)$ versus $k\eta$ for polymeric and elastic turbulence. We again see a good agreement between the scalings of PT and ET for $k\eta \gg 1$. The theoretical analysis by Fouxon and Lebedev \cite{fouxon2003spectra} predicts $E(k) \sim k^{- \alpha }$ with $\alpha >3$ and $E_p(k) \sim k^2 E(k)$. While our result for $E(k) \sim k^{-4}$ is consistent with the predicted theoretical constraint, the scaling for $E_p(k)$ does not follow the theoretical prediction.

\section{$\Sigma_2$ for PT and NT}
\begin{figure}[hbt!]
		\includegraphics[scale=0.55]{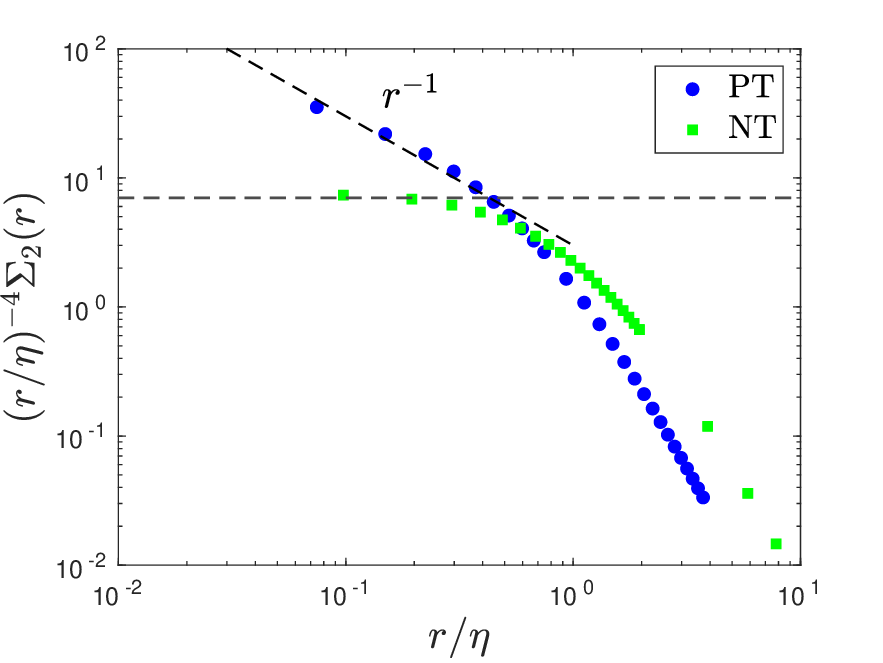}
	\caption{The comparison between PT and NT (with $Re_{\lambda} \approx 240$) of the scaling for $r/\eta \ll 1$ of the compensated second structure function of three point velocity difference $(r/\eta)^{-4} \Sigma_2(r)$.}
    \label{supp:str}
\end{figure}

Figure \ref{supp:str} shows the compensated second order structure function of the three point velocity difference ($(r/\eta)^{-4} \Sigma_2(r)$) for PT and NT, where we can see a clear difference for the two cases for $r/\eta \ll 1$. For NT, $\Sigma_2(r) \sim r^4$ in accordance with the exponential decay of the energy spectrum and the analyticity of the velocity field, while for PT $\Sigma_2(r) \sim r^3$. Note that to sufficiently resolve the $\Sigma_2$ scaling at the small scales, Newtonian turbulence results for $Re_\lambda \approx 240$ are used.

\section{Local Deborah and Reynolds number for elastic turbulence}
\begin{figure}[hbt!]
		\includegraphics[scale=0.55]{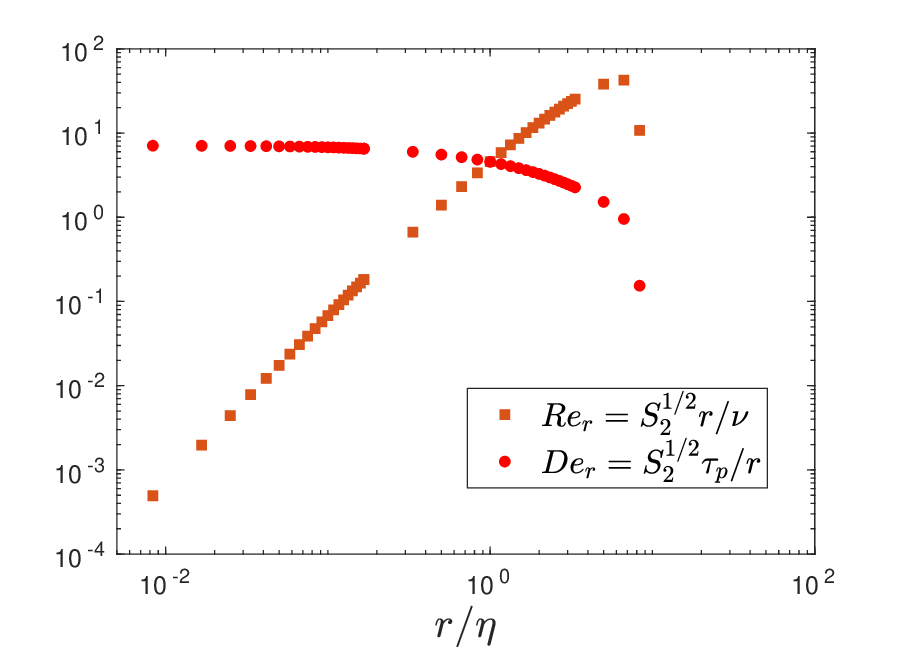}
	\caption{The variation of the local Reynolds $Re_r$ and Deborah $De_r$ numbers, versus the non-dimensional distance $r/ \eta$, for elastic turbulence with $Re_\lambda \approx 40$ and $De \approx 9$.}
    \label{supp:local_et}
\end{figure}

Figure \ref{supp:local_et} shows the variation of the local $Re_r$ and $De_r$ for the case of elastic turbulence at $Re_\lambda \approx 40$ and $De \approx 9$. For $r/\eta \ll 1$, we see that the local $De_r$ is again constant, similar to the case at large $Re_\lambda$ discussed in the main text. In elastic turbulence, the local Deborah number $De_r$ is thus both larger than the local $Re_r$ for most of the domain, as well as larger than 1. This naturally suggests the hypothesis that elastic turbulence occurs whenever $De_r >1$ and $De_r \gg Re_r$ - conditions which can be achieved even at large $Re_\lambda$ for a wide range of global $De$ as discussed in the main text and also shown in the next section.   

\section{Results at different Deborah numbers}
\begin{figure}[hbt!]
		\includegraphics[scale=0.55]{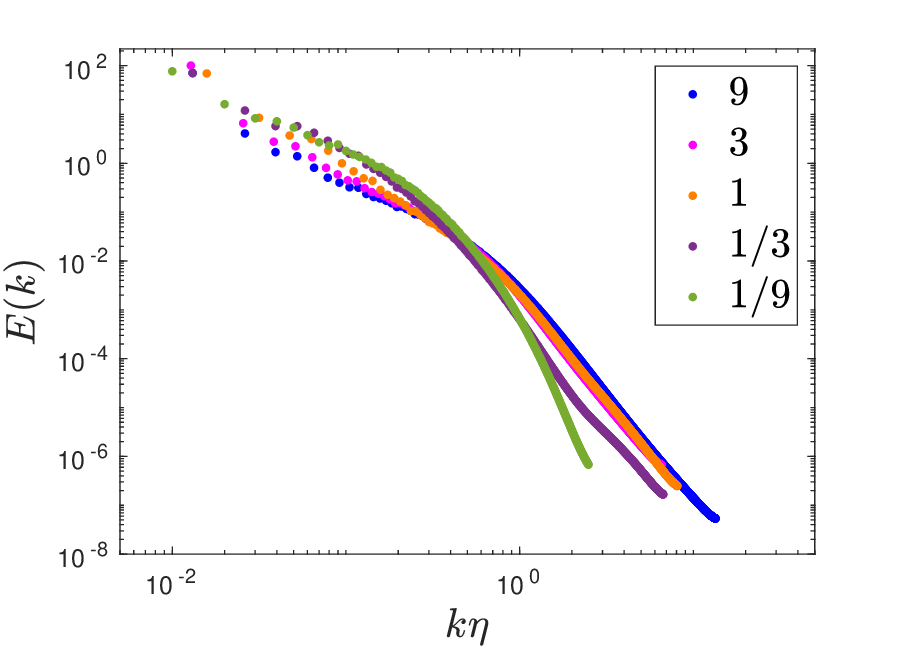}
	\caption{The variation of the energy spectrum $E(k)$ versus the non-dimensional wavenumber $k \eta$ for polymeric turbulence with $Re_\lambda \approx 240$ and varying Deborah number $De$.}
    \label{supp:spectra}
\end{figure}

\begin{figure}[hbt!]
		\includegraphics[scale=0.55]{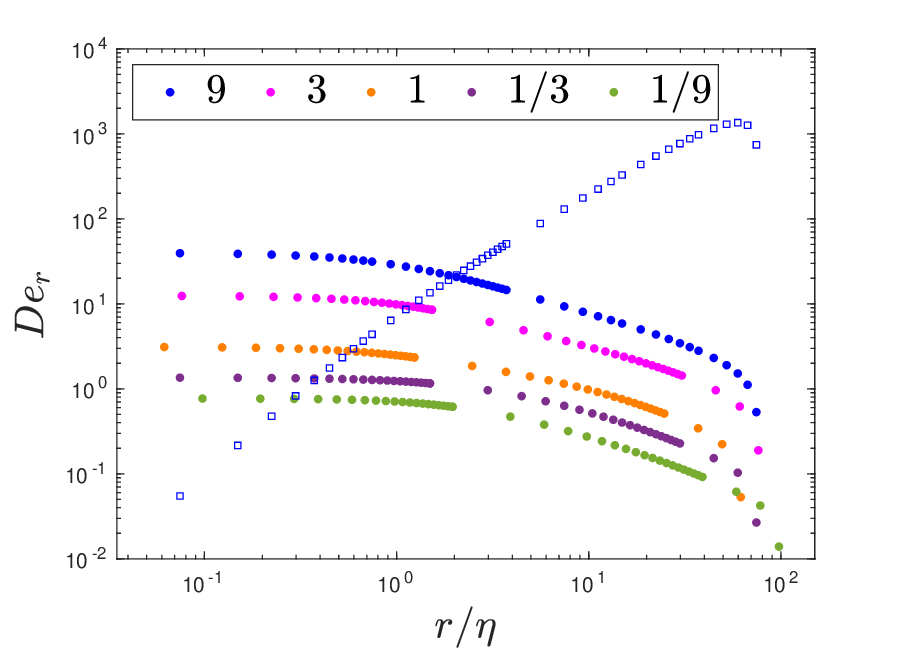}
	\caption{The variation of the local Deborah number $De_r$ versus the non-dimensional distance $r/ \eta$ for polymeric turbulence with $Re_\lambda \approx 240$ and varying Deborah number $De$. The local $Re_r$ for $De \approx 9$ is plotted as square markers to help indicate when inertia overtakes elasticity; the local $Re_r$ is mostly invariant with respect to $De$ especially for $r/\eta \ll 1$ so it is only plotted for one $De$ value for visual clarity.}
    \label{supp:local}
\end{figure}

To illustrate when elastic turbulence is seen at the small scales, here we briefly discuss results for multiple Deborah numbers. Recall that in the paper we have presented results for $Re_\lambda \approx 240$ and $De \approx 9$; here we maintain the same $Re_\lambda$ and additionally examine $De \approx 1/9, 1/3, 1, 3$. Figure \ref{supp:spectra} shows the energy spectrum for different $De$. For $k\eta > 1$, the energy spectrum exponentially decays for $De \approx 1/9$, whereas for all other cases we observe the $k^{-4}$ scaling associated with elastic turbulence. Furthermore, the energy spectrum in the dissipation range overlaps for $De \approx 1$ and above, again suggestive of the universality of elastic turbulence, where the dynamics ultimately becomes independent of $De$ at large enough $De$. Figure \ref{supp:local} shows the local Deborah number $De_r = \tau_p (S_2)^{1/2}/r$ for the various cases. For $r/\eta \ll 1$, the local $De_r$ is seen to be less than 1 for $De=1/9$, while it is greater than 1 for all the other cases; we can also note that the local Deborah number eventually becomes larger than the local Reynolds number at the small scales, confirming the hypothesis that elastic turbulence exists at scales whenever the $De_r > 1$, provided that $De_r \gg Re_r$.     

\bibliography{refer}